\pdfoutput=1
\documentclass[letterpaper,twocolumn,10pt]{article}
\usepackage[letterpaper,top=1in,bottom=1in,left=0.75in,right=0.75in,columnsep=0.33in]{geometry}
\usepackage[T1]{fontenc}
\usepackage{amsmath}
\usepackage{newtxtext,newtxmath}
\usepackage{microtype}
\usepackage{graphicx}
\usepackage{booktabs}
\usepackage{xcolor}
\usepackage{enumitem}
\usepackage{pifont}
\usepackage{tikz}
\usetikzlibrary{positioning,arrows.meta,calc,fit,backgrounds}
\usepackage[square,numbers,sort&compress]{natbib}
\PassOptionsToPackage{hyphens}{url}
\usepackage{hyperref}
\hypersetup{colorlinks=true,linkcolor=green!50!black,citecolor=red!70!black,urlcolor=blue!70!black,
  pdftitle={SplitVerify: CPU-GPU Speculative Verification for Exact Long-Context Decoding},
  pdfauthor={Aswin Raj R},
  pdfkeywords={LLM inference, KV cache offloading, speculative decoding, CPU attention}}

\definecolor{cpugreen}{HTML}{55A868}
\definecolor{pcieorange}{HTML}{DD8452}
\usepackage{etoolbox}
\AtBeginEnvironment{table}{\setlength{\belowcaptionskip}{6pt}}
\newcommand{\sys}{SplitVerify}
\newcommand{\stream}{SD+Offload}
\newcommand{\cpuonly}{CPU-only}
\newcommand{\full}{Full-KV}
\newcommand{\x}{\ensuremath{\times}}
\newcommand{\lse}{\mathrm{lse}}
\newcommand{\para}[1]{\smallskip\noindent\textbf{#1}}
\newcommand{\cmark}{\ding{51}}
\newcommand{\xmark}{\ding{55}}
\setlist{nosep,leftmargin=*}

\begin{document}

\date{}

\title{\Large\bfseries SplitVerify: CPU-GPU Speculative Verification for Exact Long-Context Decoding}

\author{Aswin Raj R\\Microsoft\\\texttt{aswinrajr@microsoft.com}}
\maketitle

\begin{abstract}

When a long context's KV cache exceeds GPU memory capacity, it must be offloaded to host memory and transferred over PCIe, which often becomes the performance bottleneck. Speculative decoding amortizes this transfer cost across multiple drafted tokens. However, some existing systems either transfer the entire offloaded KV cache to the GPU for each verification step or perform approximate attention on the CPU. The former leaves the CPU underutilized, despite its ability to read KV cache 6--12$\times$ faster than PCIe, while the latter compromises model accuracy. We observe that speculative decoding's verification phase creates an opportunity for efficient CPU attention by increasing memory reuse: verification processes $\gamma+1$ tokens at a time, allowing the CPU to reuse each fetched KV block across multiple query tokens.

In this paper, we present \textbf{\sys{}}, which partitions each layer's verification attention between the host CPU and GPU, allowing the CPU to attend to a portion of the offloaded KV cache in place. We develop a simple cost model to determine when CPU-GPU partitioning is beneficial and how to optimally divide attention computation between the two processors. Since \sys{} modifies only the verification phase, it is compatible with any drafter and preserves exactness by computing full attention over the entire KV cache. On A100 and H100 servers with 128K-token contexts, \sys{} achieves up to 2.5$\times$ higher decoding throughput than our reimplementation of SpecStream and up to 4.3$\times$ higher throughput than TriForce, using TriForce's own evaluated model.

\end{abstract}
\section{Introduction}
\label{sec:intro}

A large language model generates text one token at a time. For each new token, it computes attention over the keys and values of all earlier tokens, which it stores in a \emph{KV cache}. As the context length increases, the size of the KV cache outgrows the GPU memory capacity. For example, Llama-3.1-8B~\cite{llama3} stores 128\,KiB per token, so a 128K-token request needs 16\,GiB of cache on top of 15\,GiB of weights. Similarly, Qwen3-32B~\cite{qwen3} needs 32\,GiB of cache next to 61\,GiB of weights, more than an 80\,GB H100 can hold. The usual solution is to offload the KV cache to host memory~\cite{flexgen,infinigen,shadowkv}.

Offloading moves the bottleneck to PCIe, because the GPU must copy the offloaded cache over PCIe to decode each token. With plain offloading, Llama-3.1-8B at 128K context decodes only 1.2--1.6 tokens/s on an A100 GPU. Prior systems take one of two approaches. Some \emph{approximate} attention by reading only a selected part of the cache~\cite{infinigen,quest,shadowkv,retrievalattn,magicpig}, which is fast but loses accuracy. Others use speculative decoding~\cite{leviathan}: a cheap \emph{drafter} proposes $\gamma$ tokens and the target model checks them all in one \emph{verification} pass over $\gamma{+}1$ positions. One read of the cache then serves several tokens, and the output is exactly what the target model would produce on its own.

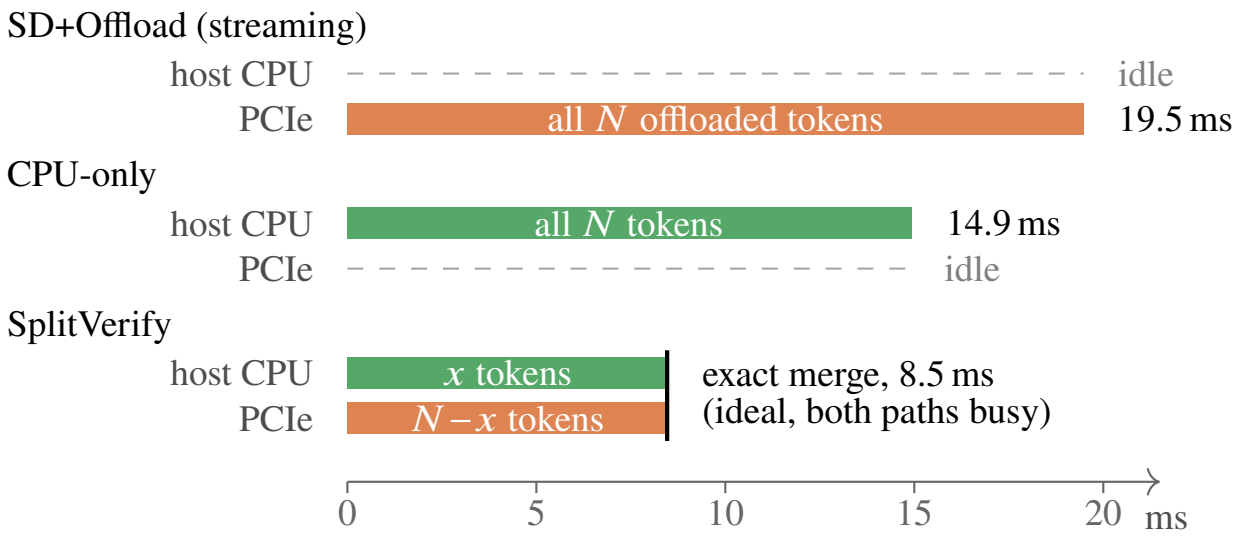
\begin{figure}[t]
\centering
\begin{tikzpicture}[x=0.25cm,y=0.3cm,font=\scriptsize,
  cpu/.style={fill=cpugreen,draw=none},pcie/.style={fill=pcieorange,draw=none},
  idle/.style={draw=black!35,dashed,thin},lab/.style={anchor=east,text=black!70}]
\foreach \y/\name in {0/SD+Offload (streaming), -3.3/CPU-only, -6.6/SplitVerify} {\node[anchor=west] at (-9.5,\y) {\name};}
\node[lab] at (-0.4,-1) {host CPU}; \draw[idle] (0,-1)--(19.48,-1); \node[anchor=west,text=black!50] at (19.9,-1) {idle};
\node[lab] at (-0.4,-2) {PCIe};     \fill[pcie] (0,-2.35) rectangle (19.48,-1.65); \node[anchor=west] at (19.9,-2) {19.5\,ms};
\node[text=white] at (9.74,-2) {all $N$ offloaded tokens};
\node[lab] at (-0.4,-4.3) {host CPU}; \fill[cpu] (0,-4.65) rectangle (14.94,-3.95); \node[anchor=west] at (15.35,-4.3) {14.9\,ms};
\node[text=white] at (7.45,-4.3) {all $N$ tokens};
\node[lab] at (-0.4,-5.3) {PCIe}; \draw[idle] (0,-5.3)--(14.89,-5.3); \node[anchor=west,text=black!50] at (15.3,-5.3) {idle};
\node[lab] at (-0.4,-7.6) {host CPU}; \fill[cpu] (0,-7.95) rectangle (8.46,-7.25); \node[text=white] at (4.23,-7.6) {$x$ tokens};
\node[lab] at (-0.4,-8.6) {PCIe};     \fill[pcie] (0,-8.95) rectangle (8.46,-8.25); \node[text=white] at (4.23,-8.6) {$N{-}x$ tokens};
\draw[thick] (8.46,-7.1)--(8.46,-9.1); \node[anchor=west,align=left] at (8.9,-8.1) {exact merge, 8.5\,ms\\[-1pt](ideal, both paths busy)};
\draw[->,black!60] (0,-10)--(21.5,-10); \node[anchor=north west,text=black!70] at (20.6,-10.25) {ms};
\foreach \t in {0,5,10,15,20} {\draw[black!60] (\t,-10)--(\t,-10.25) node[below,inner sep=1pt] {\t};}
\end{tikzpicture}
\vspace{-0.6em}
\caption{Splitting each layer's attention keeps both the CPU and PCIe busy.
Streaming leaves the CPU idle and CPU-only leaves PCIe idle. \sys{}
gives each path part of the tokens, so both finish together, and merges the
two results exactly. One verification layer of Llama-3.1-8B at 128K context
on the A100 server (24 cores, $\gamma{=}8$). The CPU time is measured, the PCIe time is the layer's KV at the measured 26.7\,GB/s, and the split time is the ideal from \S\ref{sec:model}.}
\label{fig:teaser}
\vspace{-1em}
\end{figure}

Exact systems such as TriForce~\cite{triforce} and the concurrent SpecStream~\cite{specstream} still copy the whole offloaded cache to the GPU in every verification round. While SpecStream hides part of that copy behind attention, it still transfers every byte across the slow PCIe interface. In contrast, the host CPU is capable of reading KV cache 6--12\x{} faster than PCIe can move it to the GPU. We ask whether the CPU can take over part of verification, where the KV cache already lives.

\begin{table}[t]
\centering\small
\setlength{\tabcolsep}{3pt}
\begin{tabular*}{\columnwidth}{@{\extracolsep{\fill}}lcccc@{}}
\toprule
 & & Spec. & CPU & CPU+GPU\\
System & Exact & decoding & attention & split\\
\midrule
ShadowKV~\cite{shadowkv} & \xmark & \xmark & \xmark & \xmark\\
MagicPIG~\cite{magicpig} & \xmark & \xmark & \cmark & \cmark\\
HGCA~\cite{hgca} & \xmark & \xmark & \cmark & \cmark\\
ScoutAttention~\cite{scout} & \xmark & \xmark & \cmark & \cmark\\
NEO~\cite{neo} & \cmark & \xmark & \cmark & \xmark\\
TriForce~\cite{triforce} & \cmark & \cmark & \xmark & \xmark\\
SpecStream~\cite{specstream} & \cmark & \cmark & \xmark & \xmark\\
Dovetail~\cite{dovetail} & \cmark & \cmark & \cmark & \xmark\\
\midrule
\textbf{\sys{}} & \cmark & \cmark & \cmark & \cmark\\
\bottomrule
\end{tabular*}
\caption{\sys{} and prior systems that keep the KV cache in host memory. \emph{CPU+GPU split}: CPU and GPU attend different parts of one request's KV in the same layer at the same time.}
\label{tab:related}
\end{table}

\para{Why verification suits the CPU.} In grouped-query attention \citep[GQA;][]{gqa}, several query heads share one KV head; Llama-3.1-8B has $G=4$ query heads per KV head. Ordinary decoding reads each KV head  to serve only $G$ query rows. It spends most of its time waiting on memory, and most of its arithmetic units sit idle. Verification puts these units to work: it reads each KV head once for all $(\gamma{+}1)G$ rows, 36 at $\gamma=8$. While the CPU stays memory-bound, the extra rows cost almost nothing, so the cost per checked token falls almost $(\gamma{+}1)$ times. Streaming also serves all positions with one copy, but the CPU reads the cache from DRAM much faster than PCIe can move it. Attending in place also keeps the cache off PCIe: only the queries and partial results cross it, about 0.3\,MB per layer at $\gamma=8$, against about 530\,MB to copy the layer's offloaded cache at 128K.

CPU compute, however, is limited. Once $(\gamma{+}1)G$ rows exceed what the cores can process at memory speed, CPU time grows with $\gamma$, whereas the cost of a copy does not depend on $\gamma$ at all. Which path wins thus depends on the draft length, the GQA factor, the PCIe generation and, most of all, the number of host cores (\S\ref{sec:motivation}, \S\ref{sec:model}). Because the two paths use different resources, they can also run at the same time on different tokens (Figure~\ref{fig:teaser}).

\para{\sys{}} leverages both CPUs and GPUs to accelerate long-context decoding. During each verification step, it uses CPU cores to attend to a portion of the offloaded tokens while streaming the remaining tokens to the GPU for parallel attention computation. The two partial results are then combined using the log-sum-exp identity, as in FlashAttention~\cite{flashattn} when aggregating results across tiles. The optimal partition between CPU and GPU depends on several factors, including $\gamma$, the context length, and the compute and memory bandwidth capabilities of both devices. We develop a simple cost model to systematically determine the best partition for attention computation. When CPU--GPU partitioning is not beneficial, \sys{} falls back to performing attention entirely on either the CPU or the GPU.

Prior work has explored CPU attention and partial-result merging in different settings. NEO~\cite{neo} and FastDecode~\cite{fastdecode} perform attention on CPUs during ordinary decoding, while HGCA~\cite{hgca} and ScoutAttention~\cite{scout} partition attention between CPU and GPU, offloading only selected tokens to the CPU and thereby introducing approximation. \sys{} takes a different approach: it computes exact attention over the entire offloaded range, exploits the multi-query structure of verification to make CPU attention compute-efficient, and balances CPU computation against PCIe streaming at every layer (Table~\ref{tab:related}).

\para{Results.} At 128K context, \sys{} decodes up to 2.5\x{} faster than an equally tuned streaming baseline and our reimplementation of SpecStream, and 4.3\x{} faster than TriForce's official implementation on TriForce's own model, while matching full-KV accuracy on RULER. With all host cores, most of this gain comes from verifying on the CPU. On the A100 server, \sys{} keeps almost every token on the CPU and is as fast as \cpuonly{}. The split matters when cores are scarce. With one GPU's share of the cores, \cpuonly{} is up to 1.81\x{} slower than streaming, whereas \sys{} moves work to PCIe and is as fast as streaming or faster. On the A100 server at this budget, where the two paths take similar time, \sys{} is 1.53\x{} faster than streaming and 1.49\x{} faster than \cpuonly{}, with drafting overlapped in both. Our model predicts that the gain shrinks with faster PCIe and larger GQA groups, and grows with more host compute.

\noindent Our contributions are:
\begin{itemize}
\item We show that speculative verification makes CPU attention over a host-resident cache competitive with PCIe streaming, and give a simple model that predicts which is faster (\S\ref{sec:motivation}, \S\ref{sec:model}).
\item We design and implement \sys{}, which splits each layer's offloaded cache between CPU attention and PCIe streaming, merges the results exactly, and balances the split at run time (\S\ref{sec:design}, \S\ref{sec:impl}).
\item We evaluate \sys{} on two servers and four models, including Llama-3.1-70B on two GPUs, and at batch sizes up to eight, against TriForce's official code and our reimplementations of TriForce and SpecStream, each at its best drafter and draft length (\S\ref{sec:eval}). We also report where \sys{} does not help.
\end{itemize}

\section{Background}
\label{sec:background}

\para{Offloaded KV cache.} A model with $L$ layers and $H_{\mathrm{kv}}$ KV heads of dimension $d$ stores $2LH_{\mathrm{kv}}ds$ bytes per token at $s$ bytes per element, and with grouped-query attention, $G$ query heads share each KV head. When the KV cache is offloaded to host DRAM, the PCIe bandwidth bounds decoding throughput. On our servers, host-to-device bandwidth is 26.7\,GB/s (PCIe Gen4) and 56.7\,GB/s (PCIe Gen5), compared to 329--344\,GB/s host DRAM bandwidth (Table~\ref{tab:platforms}).

\para{Speculative decoding.} A drafter proposes $\gamma$ tokens, and the target model processes all $\gamma{+}1$ positions in one pass. Under greedy decoding it accepts the longest matching prefix plus one token of its own~\cite{leviathan}. We call one draft-then-verify cycle a \emph{round}. If $t_d$ is the time to draft a token and $t_v(\gamma)$ is the time taken for verification in a round, the total time taken during the round is $\gamma t_d + t_v(\gamma)$. If the round commits $\tau(\gamma)$ tokens, throughput is $\tau/(\gamma t_d + t_v)$. At long context the drafter would need a long cache of its own, so TriForce~\cite{triforce} and MagicDec~\cite{magicdec} give it a small compressed cache instead. The drafter is either the target model itself \citep[\emph{self-speculation};][]{selfspec} or a small model.

\para{Exact attention over partitioned keys.} If disjoint key subsets give partial outputs $o_i$ with log-sum-exp normalizers $\lse_i$, the full attention output is
\begin{equation}
\lse=\log\textstyle\sum_i e^{\lse_i},\qquad o=\sum_i e^{\lse_i-\lse}\,o_i .
\label{eq:merge}
\end{equation}
This identity underlies the online softmax~\cite{onlinesoftmax}, FlashAttention~\cite{flashattn,flashattn2} and split-KV kernels~\cite{flashinfer}. The merge is exact and \sys{} uses it to combine the partial attention output of CPU and GPU.
\section{Motivation}
\label{sec:motivation}

\begin{figure}[t]
\centering
\includegraphics[width=\columnwidth]{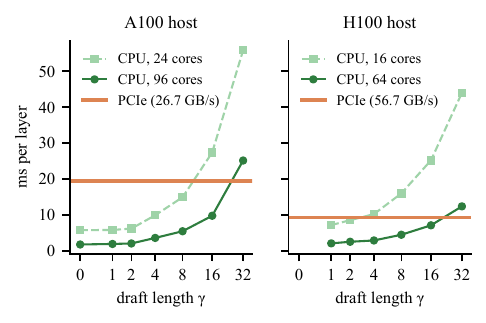}
\vspace{-1.8em}
\caption{Time to attend one layer's offloaded KV cache at 128K context (Llama-3.1-8B) during verification. Streaming time does not depend on $\gamma$, while CPU time stays low for short drafts and then grows.}
\label{fig:motivation}
\vspace{-1em}
\end{figure}

There are two exact ways to attend one layer's offloaded KV during verification: copy it to the GPU, or attend it in place on the CPU. Figure~\ref{fig:motivation} shows that neither is always faster.

\para{Streaming latency is independent of $\gamma$.} For Llama-3.1-8B at 128K context, one layer's offloaded KV cache occupies about 530\,MB in bf16. Copying it over PCIe takes about 20\,ms (Gen4) or 9.4\,ms (Gen5), however many positions the round verifies.

\para{CPU attention grows with $\gamma$.} Our kernel processes all $(\gamma{+}1)G$ query rows of a KV head in one pass over its keys and values. Up to $\gamma=2$ the kernel is memory-bound, and one layer takes about 2\,ms with all cores of either host, 9\x{} faster than the Gen4 transfer and 4\x{} faster than Gen5. Beyond that it becomes compute-bound and its time grows with $\gamma$.

\para{Neither path dominates.} On the A100 server (PCIe Gen4; 24 host cores), CPU attention becomes slower than PCIe streaming beyond $\gamma=11$. On the H100 server (PCIe Gen5; 16 host cores), this happens beyond $\gamma=3$. Each path also leaves the other's resource idle, so running both on disjoint parts of the cache can ideally add their throughput. In Figure~\ref{fig:teaser}, an ideal split would cut the layer's attention time to 8.5\,ms from 14.9\,ms on the CPU alone and 19.5\,ms with PCIe streaming.
\section{Design}
\label{sec:design}

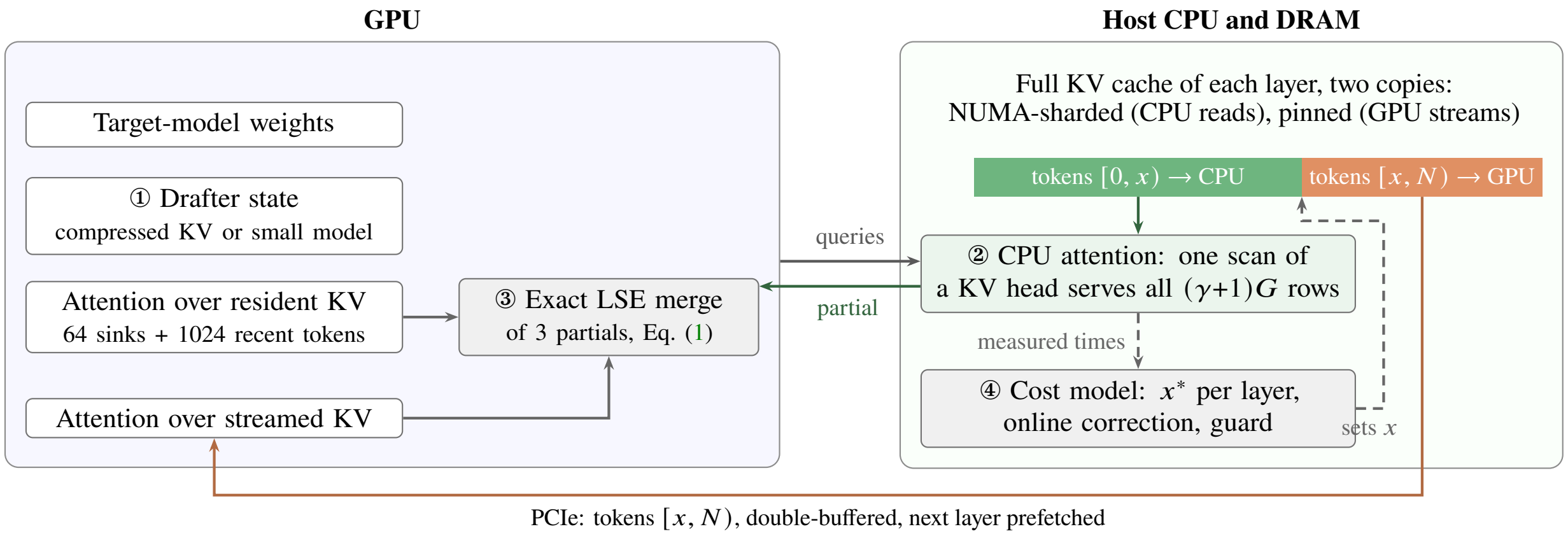
\begin{figure*}[t]
\centering
\begin{tikzpicture}[font=\footnotesize,
  box/.style={draw=black!55,rounded corners=2pt,align=center,inner sep=3pt,fill=white},
  frame/.style={draw=black!45,rounded corners=4pt,inner sep=6pt},
  arr/.style={-{Stealth[length=4pt]},thick}]
\node[box,text width=3.7cm] (w) at (2.0,2.3)  {Target-model weights};
\node[box,text width=3.7cm] (d) at (2.0,1.35) {\ding{172} Drafter state\\{\scriptsize compressed KV or small model}};
\node[box,text width=3.7cm] (r) at (2.0,0.3)  {Attention over resident KV\\{\scriptsize 64 sinks + 1024 recent tokens}};
\node[box,text width=3.7cm] (s) at (2.0,-0.75) {Attention over streamed KV};
\node[box,text width=2.9cm,fill=black!6] (m) at (6.1,0.3) {\ding{174} Exact LSE merge\\{\scriptsize of 3 partials, Eq.~(\ref{eq:merge})}};
\begin{scope}[on background layer]
\node[frame,fill=blue!3,fit={(w)(d)(r)(s)(m)(2.0,2.95)(2.0,-1.05)},label={[font=\footnotesize\bfseries]above:GPU}] (gpu) {};
\end{scope}
\node[align=center] (kvlab) at (12.6,2.55) {\footnotesize Full KV cache of each layer, two copies:\\[-1pt]\footnotesize NUMA-sharded (CPU reads), pinned (GPU streams)};
\fill[cpugreen!85]   (9.9,1.55) rectangle (13.3,1.95); \node[text=white] at (11.6,1.75) {\scriptsize tokens $[0,x)$ $\rightarrow$ CPU};
\fill[pcieorange!90] (13.3,1.55) rectangle (15.8,1.95); \node[text=white] at (14.55,1.75) {\scriptsize tokens $[x,N)$ $\rightarrow$ GPU};
\node[box,text width=4.3cm,fill=cpugreen!12] (c) at (11.6,0.75) {\ding{173} CPU attention: one scan of\\a KV head serves all $(\gamma{+}1)G$ rows};
\node[box,text width=4.3cm,fill=black!6] (cm) at (11.6,-0.65) {\ding{175} Cost model: $x^\ast$ per layer,\\online correction, guard};
\begin{scope}[on background layer]
\node[frame,fill=green!2,fit={(kvlab)(c)(cm)(15.8,1.55)(11.6,2.95)(11.6,-1.05)},label={[font=\footnotesize\bfseries]above:Host CPU and DRAM}] (host) {};
\end{scope}
\draw[arr,cpugreen!60!black] (11.6,1.55) -- (c.north);
\draw[arr,pcieorange!80!black] (14.55,1.55) -- (14.55,-1.55) -- node[below,font=\scriptsize,text=black] {PCIe: tokens $[x,N)$, double-buffered, next layer prefetched} (2.0,-1.55) -- (s.south);
\coordinate (qa) at ($(c.west)+(0,0.13)$); \coordinate (pa) at ($(c.west)+(0,-0.13)$);
\draw[arr,black!65] (gpu.east|-qa) -- node[above,font=\scriptsize] {queries} (qa);
\draw[arr,cpugreen!60!black] (pa) -- node[below,font=\scriptsize,pos=0.45] {partial} (m.east|-pa);
\draw[arr,black!60] (r.east) -- (m.west|-r.east);
\draw[arr,black!60] (s.east) -| (m.south);
\draw[arr,black!60,densely dashed] (c.south) -- node[left,font=\scriptsize] {measured times} (cm.north);
\draw[arr,black!60,densely dashed] (cm.east) -- node[below,font=\scriptsize] {sets $x$} (14.15,-0.65) -- (14.15,1.3) -- (13.3,1.3) -- (13.3,1.55);
\end{tikzpicture}
\vspace{-0.4em}
\caption{\sys{} overview. The host holds two copies of the KV cache: one sharded across NUMA nodes for CPU workers, one pinned for streaming. \ding{172}~The drafter proposes $\gamma$ tokens from GPU-resident state. \ding{173}~In each verification layer, host cores attend the first $x$ offloaded tokens while the rest are streamed to the GPU and attended there. \ding{174}~The GPU merges three partial attentions (resident, streamed, CPU). \ding{175}~A cost model sets $x$ per layer from fitted slopes and measured times.}
\label{fig:arch}
\vspace{-0.8em}
\end{figure*}

Figure~\ref{fig:arch} gives an overview. Each round drafts $\gamma$ tokens on the GPU and then verifies all $\gamma{+}1$ positions with exact attention over the whole context. Within every verification layer, the CPU and PCIe work in parallel on disjoint tokens, and the GPU merges their results.

\para{KV placement.} The full KV cache (after rotary embedding) lives in host memory in two copies: a pinned copy that the GPU streams from, and a copy split by KV head across the memory (NUMA) nodes, so that each CPU worker reads only local memory. The second copy doubles the cache's host footprint (to 32\,GiB for Llama-3.1-8B at 128K). The GPU holds the weights, the drafter's state and a small \emph{resident window}: the first 64 tokens (an attention sink~\cite{streamingllm}, which common long-context caches also keep) and the most recent 1024 tokens, whose KV was just computed on the GPU. Together they are under 1\% of a 128K context and are attended on the GPU and merged with the two other parts. Only the KV of accepted tokens is appended to the host cache.

\para{Drafting.} We use the two long-context drafters of MagicDec~\cite{magicdec}: \emph{self-speculation}, in which the target model drafts from a SnapKV-compressed~\cite{snapkv} copy of its own cache, and a \emph{small draft model} of the same family \citep[Llama-3.2-1B;][]{llama32} with its own SnapKV cache. Both caches hold 2K--4K tokens (\S\ref{sec:eval-setup}), so the cost of a draft step does not depend on the total context length of the request.

\para{CPU attention.} We wrote our own CPU attention kernel, with one version tuned for each server. On the A100 server, whose Zen~3 CPUs lack AVX-512, the kernel widens keys and values to fp32 in registers and uses AVX2 fused multiply-adds. On the H100 server (Zen~4), it uses AVX-512 to multiply queries and keys directly in bf16, accumulating in fp32. Both versions share the key idea: read each key and value only once and process all $(\gamma{+}1)G$ query rows that share a head. In each layer, \sys{} copies the queries of the $\gamma{+}1$ positions (about 150\,KB at $\gamma=8$) to the host, where the host cores attend them to the first $x$ offloaded tokens,  $[0,x)$. The cache stays in bf16 in host memory. Each core takes a range of tokens, and when it finishes, takes over work from slower cores. This makes sure one delayed core does not stall the layer. The cores' partial results are merged with Eq.~\eqref{eq:merge}. With 24 cores and $G=4$ on the A100 server, the AVX2 version is 1.3--2.8\x{} faster than the fastest PyTorch CPU attention we tried. The GPU reads the small CPU result directly from host memory, because an explicit copy would wait behind the large KV transfer on the GPU's copy engine.

\para{Split and merge.} While the CPU works on tokens $[0,x)$, the remaining tokens $[x,N)$ are copied to the GPU. The copies use  two alternating buffers, so that the next layer's copy overlaps this layer's compute. The GPU thread first enqueues all of the layer's GPU work and only then waits for the CPU, so the GPU and CPU run at the same time. The layer has three partial results from the resident tokens, the streamed tokens and the CPU, which are then merged exactly using FlashInfer's \texttt{merge\_states}~\cite{flashinfer}.

\para{Choosing the split.} The CPU part and the streamed part run at the same time, so a layer is done only when the slower of the two is done. The best split therefore gives both parts the same amount of time. To find it, we predict each part's time from its number of tokens by fitting a straight line. The CPU attends $x$ tokens and takes $t_{\mathrm{cpu}}=a_c(\gamma)\,x+b_c$; streaming handles the other $N-x$ tokens and takes $t_{\mathrm{str}}=a_s(N-x)+b_s$. Here $a$ is the time per token and $b$ the fixed time. The CPU does more work for each extra verified position, so only its time per token depends on $\gamma$; a copy costs the same however many positions are verified. Setting $t_{\mathrm{cpu}}=t_{\mathrm{str}}$ gives the balanced split
\begin{equation}
x^\ast=\frac{a_sN+b_s-b_c}{a_c+a_s}.
\label{eq:xstar}
\end{equation}
Because $x^\ast$ can fall outside the range, we limit it to $[0,N]$, and we round it to whole blocks of the CPU kernel.

\para{When not to split.} Splitting has a fixed cost. Each streamed part adds an overhead per layer ($o_h$) for launching the copy, producing an extra partial result and contending for memory. When one part is much faster than the other, this cost can outweigh the gain. A \emph{guard} adds $o_h$ to the predicted split time and keeps a single path unless the split is predicted to save at least 5\% and to stream at least 4096 tokens. 

\para{Calibration.} The fixed time $b$, the times per token $a$ and the overhead $o_h$ are measured once at startup for each model, core budget and $\gamma$, and reused across requests. $o_h$ is set to the median amount by which a layer's measured time exceeds the prediction over a few verification rounds at the model's split, or zero if the measured time is not higher. 

\para{Correction at run time.} At run time the two paths interfere, since a busy CPU delays copy launches and copies compete with CPU attention for memory bandwidth. \sys{} therefore re-estimates each slope after every layer. If a part attended $n$ tokens in time $t$, its measured time per token is $\hat a=(t-b)/n$, and the slope is updated as
\begin{equation*}
a \leftarrow 0.7\,a + 0.3\,\hat a .
\end{equation*}
A single noisy layer therefore moves the slope only a little.

\para{Core reservation.} Out of its $C$ cores, \sys{} reserves one for the thread that launches GPU work and leaves another free for the CUDA driver. Attention workers therefore do not delay copies or kernel launches. Without this reservation, rounds on the H100 server take 40\% longer (\S\ref{sec:eval-ablation}).

\para{Overlapping drafting with verification.} Drafting can be hidden behind verification, as in PEARL~\cite{pearl} and SpecStream~\cite{specstream}. Following PEARL, we give \sys{} and the streaming baseline the same optional overlap. While round $r$ verifies, a side stream keeps drafting $\gamma{+}1$ further tokens, assuming that all current drafts will be accepted. These early drafts are kept only if the assumption holds: round $r$ accepts all its drafts, and the target's own next token matches the first early draft. Otherwise they are discarded, so the output is the same as without overlap.

\para{Exactness.} By Eq.~\eqref{eq:merge}, \sys{} computes full attention over every cached token. Like any change of attention kernel, it adds numbers in a different order and so rounds some of them differently: the partial results are rounded to bf16 before the merge, as in FlashInfer, and the CPU kernel uses fused multiply-adds. \sys{} therefore accepts the same tokens as full-attention decoding except where rounding flips a near-tie between the top two candidates. Because the online split depends on timing, outputs are not bit-reproducible. For testing, we use two optional modes: a fixed split makes runs bit-reproducible, and fp32 partial results avoid the rounding to bf16 before the merge (\S\ref{sec:eval-exact}), at a cost of 0.5\% in round time (\S\ref{sec:eval-ablation}).

\section{When Does the Split Help?}
\label{sec:model}

The CPU path wins when the host has a lot of compute per unit of PCIe bandwidth, and the split helps most when the two paths take similar time. A roofline-style model~\cite{roofline} makes this precise.

\para{Model.} Consider one layer with $N$ offloaded tokens. Streaming copies $B=2NH_{\mathrm{kv}}ds$ bytes over PCIe. The CPU reads the same bytes from DRAM and performs $F=4(\gamma{+}1)GH_{\mathrm{kv}}dN$ FLOPs, because each element meets $(\gamma{+}1)G$ query rows, with one multiply-add for scores and one for values. In bf16 ($s{=}2$), this is $(\gamma{+}1)G$ FLOPs per byte. Hence
\begin{equation}
T_{\mathrm{str}}=\frac{B}{\mathrm{BW}_{\mathrm{pcie}}},\qquad
T_{\mathrm{cpu}}=\max\!\Big(\frac{B}{\mathrm{BW}_{\mathrm{dram}}},\frac{F}{P_{\mathrm{cpu}}}\Big),
\label{eq:model}
\end{equation}
where $\mathrm{BW}_{\mathrm{dram}}$ is the DRAM bandwidth available to the
cores in use and $P_{\mathrm{cpu}}$ is the FLOP rate our kernel achieves.

\para{Crossover.} Once compute-bound, the CPU beats streaming while 
$F/P_{\mathrm{cpu}} < B/\mathrm{BW}_{\mathrm{pcie}}$, that is, while
\begin{equation}
\gamma+1<\frac{P_{\mathrm{cpu}}}{G\cdot \mathrm{BW}_{\mathrm{pcie}}}.
\label{eq:crossover}
\end{equation}
At $\gamma=8$, our kernel achieves $P_{\mathrm{cpu}}=3.4$ and 1.3\,TFLOP/s with 96 and 24 cores on the A100 server, and 4.2 and 1.2\,TFLOP/s with 64 and 16 cores on the H100 server. Eq.~\eqref{eq:crossover} then places the crossover at $\gamma \approx 31$, 11, 18 and 4. These are close to the crossovers measured in Figure~\ref{fig:motivation}: about 26, 11, 23 and 3, interpolated between measured points. 

\para{Gain from splitting.} When both paths run at once on disjoint tokens, an ideal split finishes in $T_{\mathrm{cpu}}T_{\mathrm{str}}/(T_{\mathrm{cpu}}+T_{\mathrm{str}})$. Compared with the faster path alone, it saves the fraction $\min(T_{\mathrm{cpu}},T_{\mathrm{str}})/(T_{\mathrm{cpu}}+T_{\mathrm{str}})$: one half when the two paths take equal time, and little when one is much faster.

\para{Predictions.} More cores, slower PCIe, and smaller $G$ favor the CPU. Faster PCIe, fewer cores, and larger $G$ (Qwen3-32B and Llama-3.1-70B have $G=8$) favor streaming. The split helps most between these extremes. Without GQA ($G=1$), CPU attention stays memory-bound for much longer drafts, and its advantage approaches the ratio of DRAM to PCIe bandwidth. CPUs with bf16 matrix units (AMX), which our hosts lack, would raise $P_{\mathrm{cpu}}$ and move the crossover to longer drafts. Finally, because streaming time does not grow with $\gamma$, streaming prefers long drafts, CPU verification prefers short ones, and \sys{} lands in between (\S\ref{sec:eval-breakdown}).
\section{Implementation}
\label{sec:impl}

All systems we evaluate, except TriForce's official code and Hugging
Face's offloaded cache, run in one PyTorch decoding engine. They differ
only in how they attend the offloaded KV. The CPU attention kernel is
written in C++. Each layer's GPU work is captured as a CUDA graph. GPU
attention uses FlashInfer~\cite{flashinfer} with its FlashAttention-2
backend. CPU attention runs in a C++ thread pool whose workers are pinned
to cores on the NUMA node that holds their KV. For Llama-3.1-70B, the
engine uses tensor parallelism across two GPUs: each GPU holds half of the
attention heads and attends their offloaded KV with half of the host
cores.
\section{Evaluation}
\label{sec:eval}

We ask four questions. How fast is \sys{} compared to the strongest exact alternatives, including on larger models, at larger batch sizes and on a second workload (\S\ref{sec:eval-main}, \S\ref{sec:eval-sota})? Does it keep full-KV accuracy (\S\ref{sec:eval-exact})? Where does its time go, and how many host cores does it need (\S\ref{sec:eval-breakdown}, \S\ref{sec:eval-cores})? Which design choices matter and how accurate is its cost model (\S\ref{sec:eval-ablation}, \S\ref{sec:eval-model})?

\subsection{Setup}
\label{sec:eval-setup}

\begin{table}[t]
\centering\footnotesize
\setlength{\tabcolsep}{3pt}
\begin{tabular}{@{}lll@{}}
\toprule
 & A100 server & H100 server\\
\midrule
GPUs (PCIe, 80\,GB) & 4\x{} A100 & 4\x{} H100\\
Host-to-GPU link & Gen4, 26.7\,GB/s & Gen5, 56.7\,GB/s\\
CPU & 2\x{} EPYC 7V13 (Zen 3) & EPYC 9554 (Zen 4)\\
Cores; NUMA nodes & 96; 4 & 64; 1\\
Host memory & 866\,GiB & 1.5\,TiB\\
DRAM read bandwidth & 329\,GB/s & 344\,GB/s\\
Core budgets $C$ (share, all) & 24, 96 & 16, 64\\
\bottomrule
\end{tabular}
\caption{Details of the experimental platforms. Bandwidths are measured. Each experiment uses one GPU, except Llama-3.1-70B (two H100s connected by NVLink) and the four-request test (\S\ref{sec:eval-cores}).}
\label{tab:platforms}
\vspace{-1em}
\end{table}

\para{Servers and core budgets.} Table~\ref{tab:platforms} lists the two servers. Each server has four GPUs, so one GPU's share of the host is a quarter of its cores: 24 on the A100 server and 16 on the H100 server. We report two core budgets: the \emph{host share}, as when every GPU serves its own request, and \emph{all cores} (96 and 64), as on a server with one GPU. Llama-3.1-70B runs on two GPUs, so its host share is 32 cores. \S\ref{sec:eval-cores} varies the budget. Both servers run PyTorch 2.13, CUDA 13.0, and FlashInfer 0.6.18.

\para{Models and memory.} Our main model is Llama-3.1-8B-Instruct~\cite{llama3} (32 layers, $G=4$), run in bf16 with greedy decoding. To model a 24\,GB-class GPU, we cap the PyTorch allocator at 30\% of device memory (23.8\,GiB on both servers). At 128K our offloaded systems peak at 22.4\,GiB or less with one request and at 23.1\,GiB with eight, whereas full-KV decoding needs 34.3--35.4\,GiB. For the comparison with TriForce we use its model, LWM-Text-Chat-128K~\cite{lwm}, which has the Llama-2 architecture and no GQA ($G=1$), under the same cap. Two larger models run without the cap,
because their 128K caches do not fit on the GPU anyway. 
Qwen3-32B~\cite{qwen3} (64 layers, $G=8$) is extended to 128K with YaRN~\cite{yarn} (factor 4), as its model card recommends, and runs on one H100; with the full cache it would need 100\,GiB.
Llama-3.1-70B (80 layers, $G{=}8$) runs with tensor parallelism on two H100s, where its weights and its 40\,GiB cache do not fit together. Because offloading leaves GPU memory free for these two models, every system gets the same larger resident window (\S\ref{sec:design}): the most recent 21{,}504 tokens for Qwen3-32B and 16{,}384 for Llama-3.1-70B, the largest that keeps every system under 76\,GiB per GPU. Qwen3-8B~\cite{qwen3}, whose layers have the same attention shape as Llama-3.1-8B, appears only in two side experiments: EAGLE-3 acceptance and the exactness gate.

\para{Drafters.} We use MagicDec's two drafters (\S\ref{sec:design}). Self-speculation uses a 2049-token SnapKV cache (4096 for Qwen3-32B), the best size in a small pilot. The small drafter, Llama-3.2-1B, uses a 2049-token SnapKV cache on the A100 server and a 4096-token one on the H100 server. For Llama-3.1-8B, each system uses whichever drafter is faster for it, except in the batching study, where \stream{} uses self-speculation only. Qwen3-32B and LWM use self-speculation, and Llama-3.1-70B uses the small drafter.

\para{Workloads.} We measure throughput on PG-19~\cite{pg19} continuations, the dataset used by TriForce~\cite{triforce} and MagicDec~\cite{magicdec}, at 128K context on both servers and also at 32K and 64K on the A100 server. On PG-19, ``128K'' means a 130{,}560-token prompt (130{,}048 for LWM), and each request generates 256 tokens. We use 10 prompts per setting, except 9 at 128K for Llama-3.1-8B and Llama-3.1-70B and 16 in the batching study; the TriForce comparison and plain offloading use fewer, as their tables state. As a second workload, we use LongBench-v2~\cite{longbenchv2} with chain-of-thought answers capped at 1024 tokens: 10 questions with 96--120K-token inputs, the range used by Vegas~\cite{vegas}, and 10 with 24--32K-token inputs. We measure accuracy with RULER~\cite{ruler} at 64K and 128K.

\begin{table}[t]
\centering\small
\begin{tabular}{@{}ll@{}}
\toprule
System & Where offloaded KV is attended\\
\midrule
TriForce~\cite{triforce} (official) & GPU, streamed per round\\
TriForce (ours) & GPU, streamed per round\\
SpecStream (ours) & GPU, streamed in chunks\\
\stream{} (+ov) & GPU, streamed per round\\
Plain offloading & GPU, streamed per token\\
\cpuonly{} (+ov) & host CPU ($x{=}N$)\\
\sys{} (+ov) & host CPU and GPU, split\\
Full-KV & no offloading (exceeds the cap)\\
SnapKV & no offloading (compressed cache)\\
\bottomrule
\end{tabular}
\caption{Systems compared. All are exact except SnapKV, which serves only as an accuracy reference. All systems except TriForce run in our engine; plain offloading also runs in Hugging Face Transformers~\cite{hftransformers}. +ov: drafting overlaps verification (\S\ref{sec:design}). \S\ref{sec:eval-sota} checks both reimplementations.}
\label{tab:systems}
\end{table}

\para{Baselines} (Table~\ref{tab:systems}). Our main baseline is \textbf{\stream{}}, named after the baseline in SpecStream~\cite{specstream}. It is speculative decoding whose
verification copies the offloaded KV to the GPU layer by layer, overlapping each layer's copy with the previous layer's attention; we also call this path ``streaming''. \textbf{\stream{}+ov} also overlaps drafting with verification (\S\ref{sec:design}). \stream{} shares our engine, drafters, and memory cap, so the comparison isolates where the offloaded
KV is attended. It also runs close to the hardware limit: in a traced A100 run (128K, $\gamma{=}16$), its copies keep PCIe 99\% busy during verification, and every layer finishes within 3\% of the time needed just to copy its KV at the 26.6\,GB/s our engine achieves. No design that moves the whole offloaded cache every round can be much faster on this hardware. \textbf{Plain offloading} decodes without speculation and copies the cache for every token. \textbf{TriForce} runs unmodified from its official code, which supports only Llama-2-based models, on its own model; \S\ref{sec:eval-sota} describes our reimplementations of TriForce and
SpecStream. \textbf{\full{}} keeps the whole cache on the GPU without the cap and serves as an upper bound. We also tried EAGLE-3~\cite{eagle3} with public drafters in SGLang~\cite{sglang} and vLLM~\cite{vllm}, but at 128K it commits only 1.08--1.24 tokens per round for Llama-3.1-8B and Qwen3-8B and is slower than plain autoregressive decoding, consistent with the low
long-context acceptance reported by OWL~\cite{owl}. We therefore use MagicDec's long-context drafters.

\para{Tuning.} Every system gets its best configuration. For each system and drafter, we sweep the draft length over one shared grid, $\gamma\in\{1,2,4,8,16,32,48,64\}$, stop once throughput falls, and keep the drafter and $\gamma$ with the highest mean throughput over prompts. Overlapped variants are tuned the same way. Every selected optimum of \stream{}, \cpuonly{}, and \sys{} lies inside its sweep. TriForce and our reimplementation of it use their own grids (both choose $\gamma{=}24$), and the reimplementation's retrieval budget is limited by the memory cap (\S\ref{sec:eval-sota}). \full{} on the H100 server was run only at $\gamma{=}8$.

\begin{table*}[t]
\centering\small
\setlength{\tabcolsep}{8pt}
\begin{tabular}{lrrccccc}
\toprule
 & & & \multicolumn{2}{c}{\sys{} vs.\ \stream} & \cpuonly{} vs.\ \stream & \multicolumn{2}{c}{\sys{} vs.\ \cpuonly}\\
\cmidrule(lr){4-5}\cmidrule(lr){6-6}\cmidrule(lr){7-8}
Server & $C$ & Context & no overlap & +ov & no overlap & no overlap & +ov\\
\midrule
A100 & 24 & 32K & 1.19 & 1.19 & 0.97$^\circ$ & 1.22 & 1.10\\
A100 & 24 & 64K & 1.10$^\circ$ & 1.17$^\circ$ & 0.74 & 1.48 & 1.44\\
A100 & 24 & 128K & 1.41 & 1.53 & 0.96$^\circ$ & 1.47 & 1.49\\
\midrule
A100 & 96 & 32K & 1.68 & 1.62 & 1.72 & 0.98$^\circ$ & 0.99$^\circ$\\
A100 & 96 & 64K & 1.52 & 1.56 & 1.52 & 0.99$^\circ$ & 0.97$^\circ$\\
A100 & 96 & 128K & 2.19 & 2.50 & 2.19 & 1.00$^\circ$ & 1.00$^\circ$\\
\midrule
H100 & 16 & 128K & 1.07$^\circ$ & 1.14$^\circ$ & 0.55 & 1.94 & 2.09\\
\midrule
H100 & 64 & 128K & 1.65 & 1.87 & 1.48 & 1.12 & 1.14\\
\bottomrule
\end{tabular}
\caption{Paired speedups on Llama-3.1-8B (PG-19, 23.8\,GiB GPU memory cap), each system at its best drafter and $\gamma$. Each entry is the first system's speed divided by the second's; $C$ is the number of host cores. Geometric mean over 9--10 prompts (\sys{} and \cpuonly{}: mean of three runs); $^\circ$: 95\% CI includes 1. \emph{+ov}: both systems overlap drafting with verification.}
\label{tab:speedups}
\end{table*}

\para{Statistics.} Comparisons are paired: on each prompt, we divide one system's speed by the other's, and we report the geometric mean of these ratios with a 95\% $t$-interval and the number of prompts won. \sys{} and \cpuonly{} results are the mean of three runs; the ablations, LongBench-v2, and TriForce's own setting (Table~\ref{tab:triforce}) use one run. The Qwen3-32B and A100 LongBench-v2 results use a version of the guard without the overhead term $o_h$, which changes round time by only 0.1\% (\S\ref{sec:eval-ablation}).

\para{Metric.} We report decode throughput, as in the released code of TriForce and MagicDec~\cite{triforce,magicdec}: generated tokens divided by the total time of all decoding rounds. It excludes prefill and per-request setup (building the drafter's cache, capturing CUDA graphs, and one warm-up round). At 128K, setup takes a median of 4.3\,s per request on the
A100 server and 10.4\,s on the H100 server, and 21--25\,s more with TriForce's two-level drafter. Including these costs shrinks the gain: on the H100 server, \sys{}'s 1.87\x{} decode speedup at 128K with all cores and overlap becomes 1.42\x{} [1.26, 1.59] over the whole generation call and 1.14\x{} [1.08, 1.21] once the 34\,s prefill is added.

\subsection{Main results}
\label{sec:eval-main}

\begin{table}[t]
\centering\small
\setlength{\tabcolsep}{3pt}
\begin{tabular}{lrr}
\toprule
System & A100 server & H100 server\\
\midrule
Plain offloading (HF$^\ddagger$ / ours) & 1.17 / 1.57 & ---\\
\stream & 11.89 (16) & 21.04 (16)\\
\stream{}+ov & 12.19 (32) & 21.80 (16)\\
\cpuonly, all cores & 24.96 (2) & 29.36 (4)\\
\sys, host share & 16.29 (8) & 22.22 (16)\\
\sys{}+ov, host share & 17.47 (8) & 24.42 (16)\\
\sys, all cores & 24.92 (2) & 33.27 (8)\\
\sys{}+ov, all cores & \textbf{28.68} (8) & \textbf{38.72} (4)\\
\midrule
\full{} (uncapped) & 88.0 & 92.7\\
\bottomrule
\end{tabular}
\caption{Decode speed on Llama-3.1-8B at 128K (tokens/s, mean over 9 prompts; best $\gamma$ in parentheses). \full{} does not fit the 23.8\,GiB cap. Plain offloading: 5 prompts, timed over 64 generated tokens. $^\ddagger$Hugging Face's offloaded cache, run without the cap.}
\label{tab:abs}
\end{table}

\para{Llama-3.1-8B} (Tables~\ref{tab:speedups} and~\ref{tab:abs}). With all cores, \sys{} is up to 2.50\x{} faster than \stream{}. At 128K with overlap it reaches 28.7 tokens/s against 12.2 tokens/s on the A100 server and 38.7 against 21.8 on the H100 server, winning on every prompt. Most of this gain comes from verifying on the CPU: with all cores, \cpuonly{} is as fast as \sys{} on the A100 server, and \sys{} is only 1.12--1.14\x{} faster than it on the H100 server. With one GPU's share of cores, the result changes. \cpuonly{} is then up to 1.81\x{} \emph{slower} than streaming, while \sys{} shifts work to PCIe. It is up to 1.53\x{} faster than \stream{} on the A100 server and level with it on the H100 server (1.07--1.14\x{}, not significant).

\para{Why the split adds little with all cores.} On the A100 server with 96 cores, the non-overlapped configuration uses $\gamma{=}2$, where CPU attention takes about 2\,ms per layer against about 20\,ms to stream. An ideal split would save under 10\% (0.2\,ms), less than the fixed cost of a streamed part, so the guard keeps essentially every token on the CPU. The overlapped configuration uses $\gamma{=}8$, where CPU attention takes 5.5\,ms. Splitting then pays off, and \sys{} streams 22\% of the tokens over PCIe; on the H100 server it streams 23--32\%.

\para{Why the gain is smaller on the H100 server.} Gen5 PCIe doubles streaming bandwidth, which favors streaming (\S\ref{sec:model}). With 16 cores, the CPU is slower than Gen5 streaming beyond $\gamma{\approx}3$ (Figure~\ref{fig:motivation}), so the two systems end up close. The servers also differ in core count, CPU generation, NUMA layout, and GPU, so this is not a controlled comparison of PCIe generations.

\para{Overlap and plain offloading.} Overlapping drafting with verification gains 3--16\% at 128K (Table~\ref{tab:abs}). It helps \stream{} least (3--4\%), because early drafts are useful only when the whole previous draft is accepted, which is rare at the long drafts streaming prefers. Compared with plain offloading in our engine, \sys{}+ov is 19.3\x{} [13.7, 27.2] faster (A100 server, 128K, 5 prompts).

\para{Larger models.} On Qwen3-32B with all cores, \sys{} is 1.54\x{} [1.37, 1.73] faster than \stream, and 1.62\x{} [1.38, 1.90] with overlap (13.9 against 8.8 tokens/s), winning on every prompt. With 16 cores, the two are level (1.04\x{} and 1.15\x, not significant). The gain is smaller than for Llama-3.1-8B (1.65--1.87\x), as \S\ref{sec:model} predicts for $G=8$: each KV byte meets twice as many query rows, so the CPU runs out of compute at about half the draft length. On Llama-3.1-70B, also with $G=8$, \sys{} is 1.70\x{} [1.50, 1.93] faster than \stream{} with all 64 cores (10.2 against 6.2 tokens/s) and 1.33\x{} [1.12, 1.56] with 32 cores. With both GPUs copying at once, host-to-device bandwidth is about 35.6\,GB/s per GPU, and \stream{} keeps its copy engine 99\% busy. With 32 cores, \cpuonly{} reaches only 0.87\x{} the speed of streaming, and \sys{} is 1.52\x{} faster than it. vLLM~\cite{vllm} cannot serve this model at 128K on the same two GPUs; it limits the context to 28{,}928 tokens.

\para{Batches.} We also run 4 or 8 requests per GPU (A100 server, 128K, each system re-tuned). Batching does not raise total throughput for offloaded systems, because each request brings its own cache, so a larger batch moves proportionally more bytes. The total throughput of \stream{} drops as the batch grows, while that of \sys{} stays about the same, so its lead grows from 2.40\x{} at one request to 2.87\x{} at eight with all cores, and from 1.43\x{} to 1.62\x{} with 24.

\begin{figure}[t]
\centering
\includegraphics[width=\columnwidth]{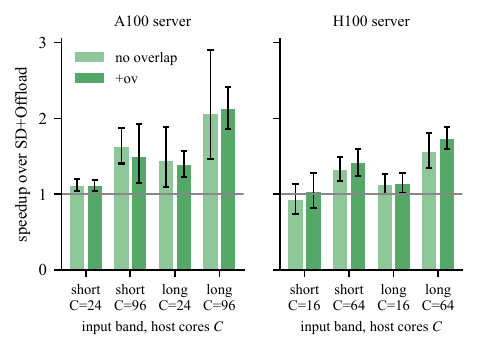}
\vspace{-1.2em}
\caption{\sys{}'s advantage carries over to LongBench-v2 and is largest on long inputs with all cores. Speedup over \stream{} (Llama-3.1-8B, chain-of-thought answers; short: 24--32K-token inputs, long: 96--120K),
with 95\% CIs.}
\label{fig:longbench}
\vspace{-1em}
\end{figure}

\para{A second workload.} LongBench-v2 tests on real questions with chain-of-thought answers of variable length (about 470--510 tokens on average), rather than fixed-length book continuations. The PG-19 results carry over (Figure~\ref{fig:longbench}). On the A100 server with 96 cores, \sys{} is 2.06\x{} [1.47, 2.90] faster than \stream{} on 96--120K-token inputs, and 2.12\x{} [1.86, 2.41] with overlap (35.5 against 17.0 tokens/s, winning on every question). On 24--32K-token inputs, it is 1.49--1.62\x{} faster. With 24 cores it is 1.39--1.44\x{} faster on long inputs and 1.11\x{} on short ones, all significant. On the H100 server, \sys{} is 1.32--1.73\x{} faster with 64 cores. With 16 cores, it is 1.13--1.14\x{} faster on long inputs and level with \stream{} on short inputs (0.92--1.02\x{}, not significant), where \stream{} uses the faster small drafter.

\subsection{Comparison with other systems}
\label{sec:eval-sota}

\begin{table}[t]
\centering\small
\setlength{\tabcolsep}{3pt}
\begin{tabular*}{\columnwidth}{@{\extracolsep{\fill}}lrrl@{}}
\toprule
System & $C$ & tok/s & vs.\ \stream{}\\
\midrule
Plain offload. & -- & 1.57 & 0.14\x{} [0.11, 0.18]\\
\stream{} & -- & 11.89 & ---\\
\stream{}+ov & -- & 12.17 & 0.98\x{}$^\circ$ [0.85, 1.13]\\
SpecStream (ours) & -- & 11.44 & 0.89\x{}$^\circ$ [0.70, 1.12]\\
TriForce (ours) & -- & 14.95 & 1.20\x{}$^\circ$ [0.87, 1.65]\\
\midrule
\cpuonly{} & 24 & 10.88 & 0.95\x{}$^\circ$ [0.78, 1.17]\\
\cpuonly{}+ov & 24 & 11.50 & 1.01\x{}$^\circ$ [0.83, 1.21]\\
\sys{} & 24 & 15.75 & 1.36\x{} [1.22, 1.52]\\
\sys{}+ov & 24 & 17.14 & \textbf{1.48\x{}} [1.31, 1.68]\\
\midrule
\cpuonly{} & 96 & 24.25 & 2.13\x{} [1.73, 2.62]\\
\cpuonly{}+ov & 96 & 27.06 & 2.37\x{} [1.96, 2.87]\\
\sys{} & 96 & 23.96 & 2.10\x{} [1.71, 2.59]\\
\sys{}+ov & 96 & 26.63 & 2.25\x{} [2.00, 2.54]\\
\midrule
\full{} (uncapped) & -- & 87.23 & 7.62\x{} [6.28, 9.26]\\
\bottomrule
\end{tabular*}
\caption{All exact systems in one run: Llama-3.1-8B, PG-19 128K, A100 server, 23.8\,GiB cap except \full{}. Mean tokens/s and paired speedup over
\stream{} (95\% CI); $^\circ$: CI includes 1. $C$: host cores; systems
marked -- do not attend on the host.}
\label{tab:allsys}
\end{table}

\begin{table}[t]
\centering\small
\setlength{\tabcolsep}{3pt}
\begin{tabular*}{\columnwidth}{@{\extracolsep{\fill}}lrl@{}}
\toprule
System & tok/s & vs.\ TriForce\\
\midrule
Autoregressive, offloaded$^\dagger$ & 0.40 & ---\\
TriForce & 3.46 & ---\\
\stream{} & 3.02 & 0.85\x{} [0.74, 0.98]\\
\midrule
\cpuonly, 24 cores & 5.62 & 1.74\x{} [1.33, 2.27]\\
\sys, 24 cores & 6.49 & \textbf{1.92\x{}} [1.59, 2.31]\\
\midrule
\cpuonly, 96 cores & 13.95 & \textbf{4.30\x{}} [3.37, 5.50]\\
\sys, 96 cores & 13.94 & \textbf{4.30\x{}} [3.37, 5.49]\\
\bottomrule
\end{tabular*}
\caption{TriForce's own setting: LWM-Text-Chat-128K on the A100 server. Mean tokens/s and paired speedup over TriForce on six test prompts (95\% CI); each system's configuration was chosen on other prompts. $^\dagger$TriForce's offloaded autoregressive decoding (default sampling), two prompts.}
\label{tab:triforce}
\end{table}

\para{All systems in one run.} In other sections, we compare \sys{} against other systems one by one. To see all exact systems side by side, we ran them in one session on the A100 server (Llama-3.1-8B, 128K, the same nine PG-19 prompts). Each system keeps the configuration selected earlier; nothing was re-tuned. 

With 24 cores, \sys{}+ov has the highest speed (Table~\ref{tab:allsys}): 1.48\x{} that of \stream{} and 1.67\x{} that of SpecStream (ours), but its lead over TriForce (ours) is not significant (1.23\x{} [0.92, 1.65]). With 96 cores, \sys{} is 2.10\x{} faster than \stream{}, 2.37\x{} faster than SpecStream (ours) and 1.75\x{} faster than TriForce (ours), winning on every prompt, and \cpuonly{} is about as fast (\sys{} 0.99\x{} [0.98, 1.00]). SpecStream (ours) is level with \stream{} at batch size 1, for reasons we give below. Here each system uses its own drafting overlap, so \sys{} does not use SpecStream's co-drafting as in Table~\ref{tab:specstream}. The rest of this subsection covers TriForce's official code and our two reimplementations.

\para{TriForce in its own setting} (Table~\ref{tab:triforce}). On TriForce's own model and code, LWM-Text-Chat-128K on the A100 server, \sys{} is 1.92\x{} faster than TriForce with 24 cores and 4.30\x{} with 96, winning on every prompt. To give TriForce its best, we tuned it on two held-out prompts over eight configurations of draft length and retrieval budget around its published ones. The best, $\gamma=24$ with a budget of 8192 tokens, lies inside the range we tried. The next larger budget, 12{,}288 tokens, needs 25.3\,GiB, more than the 23.8\,GiB cap our systems run under. Our systems' $\gamma$ was chosen on twelve other prompts. TriForce's code supports only sampling, so we run it at temperature $10^{-37}$, which is effectively greedy.

TriForce's drafter is better than ours on this model: it commits 10.4 tokens per round against 9.4, with half the draft length, which makes TriForce 1.17\x{} faster than \stream. \sys{} nevertheless wins by a wide margin, because LWM has no GQA ($G=1$). Its cache is 4\x{} larger per token, making streaming 4\x{} slower, and each byte meets 4\x{} fewer query rows, so CPU attention stays memory-bound for longer drafts. With 96 cores, the guard keeps every offloaded token on the CPU, and \sys{} is as fast as \cpuonly{} (1.00\x{}): here the advantage over TriForce comes entirely from verifying on the CPU. With 24 cores the split matters. \sys{} attends 65\% of the offloaded tokens on the CPU and is 1.11\x{} [0.93, 1.32] faster than \cpuonly{} and 2.25\x{} [1.90, 2.67] faster than \stream. Some setup differences favor TriForce: it runs in fp16 rather than bf16, counts the tokens its last round produces beyond 256, runs without our memory cap (peak 23.3 against 21.3\,GiB) and may use all 96 cores (48 OpenMP threads).

\para{TriForce's algorithm on Llama-3.1-8B.} To compare beyond TriForce's own model, whose official code supports only Llama-2-based models, we reimplemented its algorithm in our engine. A small model with a StreamingLLM cache drafts for the target model, which in turn drafts from a retrieval cache of chunks chosen by TriForce's scoring rule, and the full cache is verified by streaming. On LWM with the official configuration ($\gamma=24$, budget 8192), our reimplementation commits 1.07\x{} [0.92, 1.25] as many tokens per round and runs 1.13\x{} [0.95, 1.33] as fast as the official code over eight prompts; neither difference is significant. TriForce's timer does not wait for queued GPU work, whereas ours does. On Llama-3.1-8B at 128K we use Llama-3.2-1B as the small model and tune the draft length and retrieval budget on held-out prompts, growing the budget until the next step no longer fits our memory cap (best: $\gamma=24$, 24{,}576 tokens). The reimplementation reaches 15.0 tokens/s, 1.20\x{} [0.87, 1.65] the speed of \stream. \sys{} with the settings of Table~\ref{tab:speedups} is 1.82\x{} [1.28, 2.59] faster than it with 96 cores, winning on every prompt, while with 24 cores the difference (1.18\x{} [0.86, 1.62]) is not significant. Given TriForce's drafter, \sys{} is 2.00\x{} [1.48, 2.70] faster than TriForce's algorithm with 96 cores and 1.34\x{} [1.01, 1.77] faster with 24.

\begin{table}[t]
\centering\small
\setlength{\tabcolsep}{3pt}
\begin{tabular}{lrccc}
\toprule
 & \multicolumn{2}{c}{SpecStream (ours)} & \multicolumn{2}{c}{\sys{} vs.\ SpecStream}\\
\cmidrule(lr){2-3}\cmidrule(lr){4-5}
Context & tok/s & vs.\ \stream{}+ov & 24 cores & 96 cores\\
\midrule
32K & 37.7 & 1.02$^\circ$ & 1.08$^\circ$ & 1.51\\
64K & 27.7 & 0.89$^\circ$ & 1.28 & 1.59\\
128K & 11.5 & 0.91$^\circ$ & \textbf{1.71} & \textbf{2.49}\\
\bottomrule
\end{tabular}
\caption{SpecStream (our reimplementation) at batch size 1 (A100 server, Llama-3.1-8B). SpecStream (ours) and \sys{} both use SpecStream's co-drafting scheduler; \stream{}+ov uses our original drafting overlap; co-drafting is 0--4\% faster for it (not significant). Speedups are paired; $^\circ$: 95\% CI includes 1.}
\label{tab:specstream}
\end{table}

\para{SpecStream.} The concurrent SpecStream~\cite{specstream} is also exact. We found no code release linked from its paper and reimplemented it in our engine. For details the paper leaves open, we followed a public SGLang codebase that implements its design~\cite{specstreamrepo}. SpecStream streams the offloaded cache in 2048-token chunks (four per transfer, two staging buffers, prefetching into the next layer) and attends each chunk as it arrives, merging partial results in fp32. A cost-based controller picks the draft length per round from $\{2,4,6,8\}$. While verification waits for transfers, the drafter generates one token at a time, taking a step only if it fits in the remaining wait. We give the same co-drafting scheduler to \sys{} and \stream.

The reimplementation passes the same exactness gate as every other system (\S\ref{sec:eval-exact}) and reproduces the direction of each ablation in SpecStream's paper at 16K. Chunked streaming is 3.8\x{} faster than restoring the whole layer before attention, prefetching adds 4\%, the width controller matches a fixed width, and co-drafting adds 21\%. Not all of the 3.8\x{} comes from early attention, because the slower path also concatenates history on the CPU and copies it in one blocking transfer. Two aspects could not be matched. SpecStream confines the drafter to a subset of GPU compute units through a driver hook that our driver does not support, so our drafter runs on a low-priority stream instead. We also run one request on one GPU rather than batched requests with tensor parallelism. Our results evaluate SpecStream's design at batch size 1 and do not reproduce its batched results.

SpecStream's controller is limited to $\gamma\le8$, while streaming prefers 16--32, and a fixed $\gamma$ from our grid was always faster (11.5 against 7.7 tokens/s at 128K); we report the faster. Even so, SpecStream (ours) is no faster than \stream{}+ov at any context (0.89--1.02\x, not significant; Table~\ref{tab:specstream}). This is expected at batch size 1: \stream{} already keeps PCIe 99\% busy, so starting attention earlier on each chunk cannot move bytes faster. \sys{} instead removes bytes from the link. With the same co-drafting it is 2.49\x{} [1.73, 3.59] faster than SpecStream (ours) at 128K with 96 cores and 1.71\x{} [1.26, 2.32] with 24, on every prompt. Co-drafting is 0--4\% faster than our original overlap for \stream{} (not significant) and 0.90--1.01\x{} as fast for \sys, whose shorter rounds leave less time to hide drafting.

\subsection{Exactness and accuracy}
\label{sec:eval-exact}

\begin{table}[t]
\centering
\small
\begin{tabular}{lcc}
\toprule
System & 64K & 128K \\
\midrule
\full{} (exact) & 91.6 & 83.9 \\
\sys{} (exact) & 91.6 & 83.9 \\
SnapKV 4K (lossy) & 85.6 & 70.4 \\
SnapKV 2K (lossy) & 80.4 & 63.8 \\
\bottomrule
\end{tabular}
\caption{RULER accuracy (\%, average of 11 tasks, Llama-3.1-8B). \sys{} matches \full{}; decoding from a compressed cache does not.}
\label{tab:ruler}
\end{table}

\sys{} computes full attention, but like any change of attention kernel, it adds numbers in a different order and so can round differently. In all 153 places where an exact system (\stream, \sys{} or \cpuonly{}; PG-19 at 64K and 128K) first departs from full-KV decoding, the top two tokens are at most 0.15 logits apart. Because \sys{}'s split depends on timing, its output can also vary between runs: on the A100 server without overlap, all three runs gave the same 256-token output on only 20--56\% of prompts, whereas \cpuonly{} always did.

\para{Logit accuracy.} We test whether \sys{}'s numerical error is larger than that of other exact kernels. We feed every system the same tokens and compare its logits at each position with a reference computed entirely in fp32. As a yardstick we use nine exact bf16 runs: three attention kernels (FlashInfer, PyTorch SDPA and an fp32-math kernel), each verifying 1, 9 or 33 positions at a time. A configuration passes if it meets the tolerance for each of four measures relative to these kernels: top-1 agreement with fp32, the size of the logit error, whether the error grows with position or split, and the number of outlier positions. 

\sys{} with fixed splits and \cpuonly{} pass at both core budgets for Llama-3.1-8B and Qwen3-8B at 32K and 128K, and for Qwen3-32B at 32K. The online split passes for Llama-3.1-8B at 128K, and all five tested configurations pass for Llama-3.1-70B with tensor parallelism at 32K. On LWM, we can test only 32K, the longest context whose fp32 reference fits on the GPU. We tested CPU-only verification and fixed splits at that length. One fixed split fails only the outlier measure, with one position in 768 (96 cores, $x/N=0.25$). With fp32 partials it has none, so the outlier comes from rounding partial results to bf16 before the merge (\S\ref{sec:design}). Our engine itself agrees with Hugging Face Transformers~\cite{hftransformers} on the top token at 255 of 256 positions; the exception is the most nearly tied one (0.09 logits).

\para{Task accuracy.} Table~\ref{tab:ruler} reports RULER on the 11 tasks used by ShadowKV~\cite{shadowkv}, scored with RULER's own metrics (96 samples per task at 64K, and 16 at 128K to bound run time). \sys{} matches \full{}: its outputs are token-identical on 98\% of samples, and the rest diverge at near-ties. Decoding directly from a SnapKV cache of the drafter's size (2K) or twice that loses 6--20 points on average and up to 94 on a single task (\texttt{niah\_single\_3} at 128K falls from 100 to 6.25). The self-speculative drafter works from such a cache, so exact verification over the whole cache is what preserves accuracy.

\subsection{Where the time goes}
\label{sec:eval-breakdown}

\begin{figure}[t]
\centering
\includegraphics[width=\columnwidth]{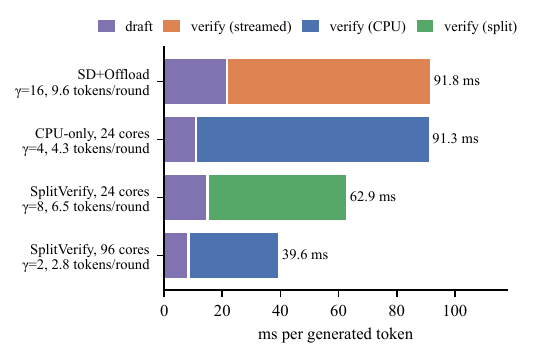}
\vspace{-1.2em}
\caption{Time per generated token at 128K on the A100 server (Llama-3.1-8B, each system at its best $\gamma$). Splitting cuts verification time when cores are few; with all cores, \sys{} keeps every token on the CPU.}
\label{fig:breakdown}
\vspace{-1em}
\end{figure}

Figure~\ref{fig:breakdown} breaks down the A100 result at 128K. \stream{} spends 70.0\,ms per generated token on verification, because each round streams the whole offloaded cache (about 20\,ms per layer), and its best $\gamma=16$ spreads that over only 9.6 tokens per round. With 24 cores, \cpuonly{} verification is slower still at 80.5\,ms, but splitting brings it down to 48.0\,ms. With 96 cores, \sys{} keeps all tokens on the CPU and verifies in 31.2\,ms. Since CPU verification cost grows with $\gamma$, \sys{} prefers short drafts ($\gamma=8$ with 24 cores and 2 with 96), which waste fewer drafted tokens, so drafting costs 14.9 and 8.4\,ms per token against 21.7\,ms for \stream.

\subsection{How many host cores are needed?}
\label{sec:eval-cores}

\begin{figure}[t]
\centering
\includegraphics[width=\columnwidth]{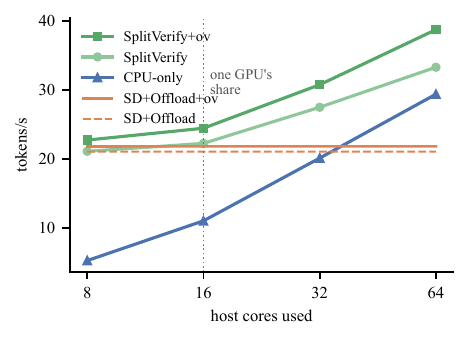}
\vspace{-1.2em}
\caption{Decode speed as the host cores given to one GPU vary (H100 server, Llama-3.1-8B, 128K, best $\gamma$ per point; \stream{} at its best $\gamma{=}16$ throughout). The vertical dotted line marks one GPU's share (16 cores). \sys{} matches streaming with few cores and pulls ahead with more.}
\label{fig:cores}
\vspace{-1em}
\end{figure}

Figure~\ref{fig:cores} varies the host cores given to one GPU on the H100 server from 8, half of one GPU's share, to all 64. Streaming does not use host cores, so its speed does not change. \cpuonly{} grows steadily from 5.2 to 29.4 tokens/s and overtakes streaming only between 32 and 64 cores. \sys{} with overlap is level with streaming at 8 and 16 cores (1.06\x{} and 1.14\x) and clearly faster at 32 and 64 (1.46\x{} [1.27, 1.68] and 1.87\x{} [1.56, 2.25]). Without overlap, \sys{} is 3.9\x{} faster than \cpuonly{} at 8 cores, because it moves work to PCIe.

\para{CPU cost.} The speedup is paid for in host CPU time. In the same runs, \sys{}+ov occupied 0.29 core-seconds per generated token at 8 cores and 1.68 at 64, against 0.05 for streaming, measured as process CPU time over the decoding rounds. Worker threads stay busy for the whole decode, so this is CPU time occupied rather than the minimum needed. \sys{} is therefore best suited to hosts whose cores are otherwise idle.

\para{Four requests at once.} The results so far run one request while the other GPUs are idle. To test contention, we ran four 128K requests at once on the A100 server, one per GPU, each confined to its own NUMA node's 24 cores (streaming too), and ran the whole schedule twice. Shared host memory slows CPU attention: \cpuonly{} loses 21\% of its isolated speed and \sys{} 13\%, while streaming is unaffected. \sys{} still beats streaming by 1.23\x{} [1.10, 1.38] (1.39\x{} with overlap), and the four GPUs together decode 57.0 tokens/s against 47.6. Its lead over \cpuonly{} grows from 1.47\x{} in isolation to 1.62\x{} [1.46, 1.80], because it moves more work to PCIe as the CPU slows.

\subsection{Ablations}
\label{sec:eval-ablation}

\begin{figure}[t]
\centering
\includegraphics[width=\columnwidth]{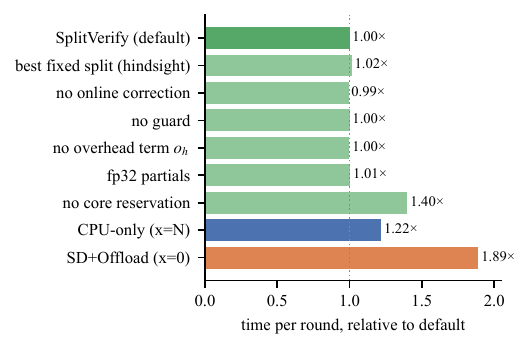}
\vspace{-1.2em}
\caption{Removing one component at a time (H100 server, Llama-3.1-8B, 128K, 64 cores, $\gamma{=}8$ for every bar, no overlap, one run). Bars show wall time per round relative to the default, geometric mean over prompts; lower is better.}
\label{fig:ablation}
\vspace{-1em}
\end{figure}

Figure~\ref{fig:ablation} removes one component at a time. Because acceptance varies between runs, we compare wall time per round, paired over prompts, rather than throughput. Core reservation matters most: without it, rounds take 40\% longer, longer even than with \cpuonly. The online split beats every fixed split we tried: its rounds are 1.5\% faster than with the best fixed split in hindsight (CPU fraction 0.7), and 3\% and 8.5\% faster than with fractions 0.6 and 0.8. Online correction, the guard and the overhead term $o_h$ each change round time by less than 1\% on this dedicated host, and fp32 partials add 0.5\%.

\subsection{Accuracy of the cost model}
\label{sec:eval-model}

What matters for speed is the split the model picks, not the precision of its time estimates. We froze the calibration and measured per-layer time at four splits in 16 settings on the A100 server spanning $\gamma\in\{2,4,8,16\}$, 32K and 128K, and both core budgets. In all 16, the split the model chooses is within 10\% of the fastest split measured, and within 1\% in 12. Streaming time is predicted within 0.3\%. CPU and split times are slightly under-predicted, by a fixed cost per layer (median 0.3--0.5\,ms) that does not grow with the number of tokens and that the guard's overhead term $o_h$ accounts for (\S\ref{sec:design}).

\section{Related Work}
\label{sec:related}

\para{Speculative decoding for long contexts.} TriForce~\cite{triforce} drafts hierarchically from a retrieval-based partial cache and a small model. MagicDec~\cite{magicdec} shows that speculation pays at long context and drafts from StreamingLLM or SnapKV caches. LongSpec~\cite{longspec} and OWL~\cite{owl} train drafters for long contexts, QuantSpec~\cite{quantspec} drafts from a quantized cache, and SparseSpec~\cite{sparsespec} drafts with sparse attention for long reasoning outputs while keeping part of the cache in host memory. \sys{} is orthogonal to the drafter and changes where verification attends an offloaded cache. The concurrent SpecStream~\cite{specstream} starts verification attention on each chunk of the host-resident cache as it arrives and overlaps drafting with it, targeting batched serving. \sys{} is complementary: it computes part of verification on the host instead of moving it, and its streamed part could adopt SpecStream's chunked streaming (\S\ref{sec:eval-sota}). Also concurrent, Adaptive DHiSparse~\cite{dhisparse} combines speculative decoding with a host-memory KV hierarchy from which selected rows are fetched. SpecExec~\cite{specexec} and Sequoia~\cite{sequoia} use speculation to amortize offloaded \emph{weights}.

\para{KV offloading and sparse attention.} FlexGen~\cite{flexgen} schedules offloaded tensors for throughput-oriented batch inference. InfiniGen~\cite{infinigen}, ShadowKV~\cite{shadowkv}, Quest~\cite{quest}, RetrievalAttention~\cite{retrievalattn}, and MagicPIG~\cite{magicpig} attend to a selected or sampled subset of the cache, which is fast but approximate; Table~\ref{tab:ruler} shows how much accuracy a compressed cache can lose. LMCache~\cite{lmcache} and HiCache~\cite{hicache} use host memory to reuse prefixes across requests.

\para{CPUs in LLM inference.} NEO~\cite{neo} runs decoding attention on the local host CPU and FastDecode~\cite{fastdecode} on CPUs across several nodes, both to free GPU memory for larger batches. APEX~\cite{apex} schedules such hybrid execution, and FlexGen~\cite{flexgen} and MoE-Lightning~\cite{moelightning} can also attend on the CPU. All of these target ordinary decoding, with one new query position per request and step. HGCA~\cite{hgca,beyondgpu} attends to recent tokens on the GPU and to selected salient tokens on the CPU and merges the two parts with the log-sum-exp identity, and ScoutAttention~\cite{scout} keeps critical blocks on the GPU while attending further selected blocks on the CPU. Because their CPU side attends only a selected subset, these hybrids are approximate. HybridGen~\cite{hybridgen} also divides attention between CPU and GPU, on hosts with CXL-expanded memory, without speculation. Dovetail~\cite{dovetail} runs the draft model on the GPU and the target model on the CPU. DirectKV~\cite{directkv} lets GPU kernels read host-resident KV directly over the NVLink-C2C link of Grace Hopper systems, and ECHO~\cite{echo} prefetches offloaded KV losslessly for models with native sparse attention, whereas we target PCIe-attached GPUs and dense attention. Fiddler~\cite{fiddler}, KTransformers~\cite{ktransformers}, and llama.cpp~\cite{llamacpp} split layers or experts between CPU and GPU, and PowerInfer~\cite{powerinfer} splits neurons. Unlike all of these, \sys{} attends the \emph{whole} offloaded cache exactly during speculative verification and divides each layer's attention between host cores and PCIe according to a cost model.
\section{Limitations and Conclusion}
\label{sec:limits}

\para{Limitations.} \sys{} targets long-context requests at small batch sizes (we test up to eight per GPU) and relies on host cores that would otherwise be idle. Keeping a second, NUMA-sharded copy of the cache doubles its host memory footprint (16\,GiB extra for Llama-3.1-8B at 128K). Our speedups are for decoding; once setup and prefill are included, the gain on a single 128K request falls to 1.14\x{} on the H100 server. Faster interconnects (PCIe Gen6, CPU--GPU links such as NVLink-C2C) and larger GQA factors shrink the advantage, while CPUs with matrix units (AMX), which our hosts lack, should grow it. With all host cores, verifying on the CPU alone is as fast as \sys{} on the A100 server and at most 12\% slower for Llama-3.1-8B on the H100 server (14\% for Llama-3.1-70B), so the split pays off mainly when cores are scarce or shared. Finally, online splitting is not bit-reproducible across runs, although a fixed-split mode is.

\para{Conclusion.} When the KV cache lives in host memory, speculative verification can attend part of it in place, because checking many drafted tokens at once makes CPU attention efficient. \sys{} splits each layer's verification attention between host cores and PCIe streaming, merges the parts exactly and chooses the split with a simple cost model. It therefore follows whichever path is faster on a given host and is faster than both when they take similar time. At 128K it decodes up to 2.5\x{} faster than exact streaming and 4.3\x{} faster than TriForce on its own model, with full-KV accuracy.

\bibliographystyle{plainnat}
\bibliography{refs}

@article{specstream,
  title={{Resource-Efficient Speculative Decoding for Long-Context LLM Serving}},
  author={Fei Li and Song Liu and Shiqiang Nie and Jinyu Wang and Weiguo Wu},
  journal={arXiv preprint arXiv:2609.33184},
  year={2026}
}

@inproceedings{dovetail,
  title={{Dovetail: A CPU/GPU Heterogeneous Speculative Decoding for LLM inference}},
  author={Libo Zhang and Zhaoning Zhang and Baizhou Xu and Rui Li and Zhiliang Tian and Songzhu Mei and Dongsheng Li},
  booktitle={Conference on Empirical Methods in Natural Language Processing (EMNLP)},
  year={2025}
}

@inproceedings{specexec,
  title={{SpecExec: Massively Parallel Speculative Decoding for Interactive LLM Inference on Consumer Devices}},
  author={Ruslan Svirschevski and Avner May and Zhuoming Chen and Beidi Chen and Zhihao Jia and Max Ryabinin},
  booktitle={Advances in Neural Information Processing Systems (NeurIPS)},
  year={2024}
}

@inproceedings{sequoia,
  title={{Sequoia: Scalable and Robust Speculative Decoding}},
  author={Zhuoming Chen and Avner May and Ruslan Svirschevski and Yuhsun Huang and Max Ryabinin and Zhihao Jia and Beidi Chen},
  booktitle={Advances in Neural Information Processing Systems (NeurIPS)},
  year={2024}
}

@article{lmcache,
  title={{LMCache: An Efficient KV Cache Layer for Enterprise-Scale LLM Inference}},
  author={Yuhan Liu and Yihua Cheng and Jiayi Yao and Yuwei An and Xiaokun Chen and Shaoting Feng and Yuyang Huang and Samuel Shen and Rui Zhang and Kuntai Du and Junchen Jiang},
  journal={arXiv preprint arXiv:2510.09665},
  year={2025}
}

@inproceedings{magicdec,
  title={{MagicDec: Breaking the Latency-Throughput Tradeoff for Long Context Generation with Speculative Decoding}},
  author={Ranajoy Sadhukhan and Jian Chen and Zhuoming Chen and Vashisth Tiwari and Ruihang Lai and Jinyuan Shi and Ian En-Hsu Yen and Avner May and Tianqi Chen and Beidi Chen},
  booktitle={International Conference on Learning Representations (ICLR)},
  year={2025}
}

@article{hgca,
  title={{HGCA: Hybrid GPU-CPU Attention for Long Context LLM Inference}},
  author={Weishu Deng and Yujie Yang and Peiran Du and Lingfeng Xiang and Zhen Lin and Chen Zhong and Song Jiang and Hui Lu and Jia Rao},
  journal={arXiv preprint arXiv:2507.03153},
  year={2025}
}

@inproceedings{scout,
  title={{ScoutAttention: Efficient KV Cache Offloading via Layer-Ahead CPU Pre-computation for LLM Inference}},
  author={Qiuyang Zhang and Kai Zhou and Ding Tang and Kai Lu and Cheng Li and Zhenyu Yang and Peng Xu and Jiguang Wan},
  booktitle={ACM/IEEE Design Automation Conference (DAC)},
  year={2026}
}

@inproceedings{neo,
  title={{NEO: Saving GPU Memory Crisis with CPU Offloading for Online LLM Inference}},
  author={Xuanlin Jiang and Yang Zhou and Shiyi Cao and Ion Stoica and Minlan Yu},
  booktitle={Conference on Machine Learning and Systems (MLSys)},
  year={2025}
}

@inproceedings{apex,
  title={{APEX: Asynchronous Parallel CPU-GPU Execution for Online LLM Inference on Constrained GPUs}},
  author={Jiakun Fan and Yanglin Zhang and Xiangchen Li and Dimitrios S. Nikolopoulos},
  booktitle={IEEE International Parallel and Distributed Processing Symposium (IPDPS)},
  year={2026}
}

@inproceedings{flexgen,
  title={{FlexGen: High-Throughput Generative Inference of Large Language Models with a Single GPU}},
  author={Ying Sheng and Lianmin Zheng and Binhang Yuan and Zhuohan Li and Max Ryabinin and Beidi Chen and Percy Liang and Christopher R{\'e} and Ion Stoica and Ce Zhang},
  booktitle={International Conference on Machine Learning (ICML)},
  year={2023}
}

@inproceedings{moelightning,
  title={{MoE-Lightning: High-Throughput MoE Inference on Memory-constrained GPUs}},
  author={Shiyi Cao and Shu Liu and Tyler Griggs and Peter Schafhalter and Xiaoxuan Liu and Ying Sheng and Joseph E. Gonzalez and Matei Zaharia and Ion Stoica},
  booktitle={ACM International Conference on Architectural Support for Programming Languages and Operating Systems (ASPLOS)},
  year={2025}
}

@article{fastdecode,
  title={{FastDecode: High-Throughput GPU-Efficient LLM Serving using Heterogeneous Pipelines}},
  author={Jiaao He and Jidong Zhai},
  journal={arXiv preprint arXiv:2403.11421},
  year={2024}
}

@inproceedings{fiddler,
  title={{Fiddler: CPU-GPU Orchestration for Fast Inference of Mixture-of-Experts Models}},
  author={Keisuke Kamahori and Tian Tang and Yile Gu and Kan Zhu and Baris Kasikci},
  booktitle={International Conference on Learning Representations (ICLR)},
  year={2025}
}

@inproceedings{powerinfer,
  title={{PowerInfer: Fast Large Language Model Serving with a Consumer-grade GPU}},
  author={Yixin Song and Zeyu Mi and Haotong Xie and Haibo Chen},
  booktitle={ACM Symposium on Operating Systems Principles (SOSP)},
  year={2024}
}

@inproceedings{shadowkv,
  title={{ShadowKV: KV Cache in Shadows for High-Throughput Long-Context LLM Inference}},
  author={Hanshi Sun and Li-Wen Chang and Wenlei Bao and Size Zheng and Ningxin Zheng and Xin Liu and Harry Dong and Yuejie Chi and Beidi Chen},
  booktitle={International Conference on Machine Learning (ICML)},
  year={2025}
}

@inproceedings{retrievalattn,
  title={{RetrievalAttention: Accelerating Long-Context LLM Inference via Vector Retrieval}},
  author={Di Liu and Meng Chen and Baotong Lu and Huiqiang Jiang and Zhenhua Han and Qianxi Zhang and Qi Chen and Chengruidong Zhang and Bailu Ding and Kai Zhang and Chen Chen and Fan Yang and Yuqing Yang and Lili Qiu},
  booktitle={Advances in Neural Information Processing Systems (NeurIPS)},
  year={2025}
}

@inproceedings{magicpig,
  title={{MagicPIG: LSH Sampling for Efficient LLM Generation}},
  author={Zhuoming Chen and Ranajoy Sadhukhan and Zihao Ye and Yang Zhou and Jianyu Zhang and Niklas Nolte and Yuandong Tian and Matthijs Douze and L{\'e}on Bottou and Zhihao Jia and Beidi Chen},
  booktitle={International Conference on Learning Representations (ICLR)},
  year={2025}
}

@inproceedings{triforce,
  title={{TriForce: Lossless Acceleration of Long Sequence Generation with Hierarchical Speculative Decoding}},
  author={Hanshi Sun and Zhuoming Chen and Xinyu Yang and Yuandong Tian and Beidi Chen},
  booktitle={Conference on Language Modeling (COLM)},
  year={2024}
}

@inproceedings{quantspec,
  title={{QuantSpec: Self-Speculative Decoding with Hierarchical Quantized KV Cache}},
  author={Rishabh Tiwari and Haocheng Xi and Aditya Tomar and Coleman Hooper and Sehoon Kim and Maxwell Horton and Mahyar Najibi and Michael W. Mahoney and Kurt Keutzer and Amir Gholami},
  booktitle={International Conference on Machine Learning (ICML)},
  year={2025}
}

@inproceedings{vegas,
  title={{Vegas: Self-Speculative Decoding with Verification-Guided Sparse Attention}},
  author={Yikang Yue and Yuqi Xue and Jian Huang},
  booktitle={International Conference on Machine Learning (ICML)},
  year={2026}
}

@inproceedings{pearl,
  title={{PEARL: Parallel Speculative Decoding with Adaptive Draft Length}},
  author={Tianyu Liu and Yun Li and Qitan Lv and Kai Liu and Jianchen Zhu and Winston Hu and Xiao Sun},
  booktitle={International Conference on Learning Representations (ICLR)},
  year={2025}
}

@inproceedings{snapkv,
  title={{SnapKV: LLM Knows What You are Looking for Before Generation}},
  author={Yuhong Li and Yingbing Huang and Bowen Yang and Bharat Venkitesh and Acyr Locatelli and Hanchen Ye and Tianle Cai and Patrick Lewis and Deming Chen},
  booktitle={Advances in Neural Information Processing Systems (NeurIPS)},
  year={2024}
}

@inproceedings{ruler,
  title={{RULER: What's the Real Context Size of Your Long-Context Language Models?}},
  author={Cheng-Ping Hsieh and Simeng Sun and Samuel Kriman and Shantanu Acharya and Dima Rekesh and Fei Jia and Boris Ginsburg},
  booktitle={Conference on Language Modeling (COLM)},
  year={2024}
}

@inproceedings{longbenchv2,
  title={{LongBench v2: Towards Deeper Understanding and Reasoning on Realistic Long-context Multitasks}},
  author={Yushi Bai and Shangqing Tu and Jiajie Zhang and Hao Peng and Xiaozhi Wang and Xin Lv and Shulin Cao and Jiazheng Xu and Lei Hou and Yuxiao Dong and Jie Tang and Juanzi Li},
  booktitle={Annual Meeting of the Association for Computational Linguistics (ACL)},
  year={2025}
}

@inproceedings{pg19,
  title={{Compressive Transformers for Long-Range Sequence Modelling}},
  author={Jack W. Rae and Anna Potapenko and Siddhant M. Jayakumar and Chloe Hillier and Timothy P. Lillicrap},
  booktitle={International Conference on Learning Representations (ICLR)},
  year={2020}
}

@inproceedings{yarn,
  title={{YaRN: Efficient Context Window Extension of Large Language Models}},
  author={Bowen Peng and Jeffrey Quesnelle and Honglu Fan and Enrico Shippole},
  booktitle={International Conference on Learning Representations (ICLR)},
  year={2024}
}

@article{qwen3,
  title={{Qwen3 Technical Report}},
  author={An Yang and Anfeng Li and Baosong Yang and Beichen Zhang and Binyuan Hui and Bo Zheng and Bowen Yu and Chang Gao and Chengen Huang and Chenxu Lv and Chujie Zheng and Dayiheng Liu and Fan Zhou and Fei Huang and Feng Hu and Hao Ge and Haoran Wei and Huan Lin and Jialong Tang and Jian Yang and Jianhong Tu and Jianwei Zhang and Jianxin Yang and Jiaxi Yang and Jing Zhou and Jingren Zhou and Junyang Lin and Kai Dang and Keqin Bao and Kexin Yang and Le Yu and Lianghao Deng and Mei Li and Mingfeng Xue and Mingze Li and Pei Zhang and Peng Wang and Qin Zhu and Rui Men and Ruize Gao and Shixuan Liu and Shuang Luo and Tianhao Li and Tianyi Tang and Wenbiao Yin and Xingzhang Ren and Xinyu Wang and Xinyu Zhang and Xuancheng Ren and Yang Fan and Yang Su and Yichang Zhang and Yinger Zhang and Yu Wan and Yuqiong Liu and Zekun Wang and Zeyu Cui and Zhenru Zhang and Zhipeng Zhou and Zihan Qiu},
  journal={arXiv preprint arXiv:2505.09388},
  year={2025}
}

@inproceedings{leviathan,
  title={{Fast Inference from Transformers via Speculative Decoding}},
  author={Yaniv Leviathan and Matan Kalman and Yossi Matias},
  booktitle={International Conference on Machine Learning (ICML)},
  year={2023}
}

@inproceedings{selfspec,
  title={{Draft \& Verify: Lossless Large Language Model Acceleration via Self-Speculative Decoding}},
  author={Jun Zhang and Jue Wang and Huan Li and Lidan Shou and Ke Chen and Gang Chen and Sharad Mehrotra},
  booktitle={Annual Meeting of the Association for Computational Linguistics (ACL)},
  year={2024}
}

@inproceedings{flashattn,
  title={{FlashAttention: Fast and Memory-Efficient Exact Attention with IO-Awareness}},
  author={Tri Dao and Daniel Y. Fu and Stefano Ermon and Atri Rudra and Christopher Ré},
  booktitle={Advances in Neural Information Processing Systems (NeurIPS)},
  year={2022}
}

@inproceedings{sglang,
  title={{SGLang: Efficient Execution of Structured Language Model Programs}},
  author={Lianmin Zheng and Liangsheng Yin and Zhiqiang Xie and Chuyue Sun and Jeff Huang and Cody Hao Yu and Shiyi Cao and Christos Kozyrakis and Ion Stoica and Joseph E. Gonzalez and Clark Barrett and Ying Sheng},
  booktitle={Advances in Neural Information Processing Systems (NeurIPS)},
  year={2024}
}

@inproceedings{vllm,
  title={{Efficient Memory Management for Large Language Model Serving with PagedAttention}},
  author={Woosuk Kwon and Zhuohan Li and Siyuan Zhuang and Ying Sheng and Lianmin Zheng and Cody Hao Yu and Joseph E. Gonzalez and Hao Zhang and Ion Stoica},
  booktitle={ACM Symposium on Operating Systems Principles (SOSP)},
  year={2023}
}

@inproceedings{flashinfer,
  title={{FlashInfer: Efficient and Customizable Attention Engine for LLM Inference Serving}},
  author={Zihao Ye and Lequn Chen and Ruihang Lai and Wuwei Lin and Yineng Zhang and Stephanie Wang and Tianqi Chen and Baris Kasikci and Vinod Grover and Arvind Krishnamurthy and Luis Ceze},
  booktitle={Conference on Machine Learning and Systems (MLSys)},
  year={2025}
}

@inproceedings{streamingllm,
  title={{Efficient Streaming Language Models with Attention Sinks}},
  author={Guangxuan Xiao and Yuandong Tian and Beidi Chen and Song Han and Mike Lewis},
  booktitle={International Conference on Learning Representations (ICLR)},
  year={2024}
}

@inproceedings{infinigen,
  title={{InfiniGen: Efficient Generative Inference of Large Language Models with Dynamic KV Cache Management}},
  author={Wonbeom Lee and Jungi Lee and Junghwan Seo and Jaewoong Sim},
  booktitle={USENIX Symposium on Operating Systems Design and Implementation (OSDI)},
  year={2024}
}

@misc{llamacpp,
  title={{llama.cpp}},
  author={Gerganov, Georgi and others},
  howpublished={\url{https://github.com/ggml-org/llama.cpp}},
  year={2026}
}

@misc{hicache,
  title={{Hierarchical KV Caching (HiCache)}},
  author={{SGLang Team}},
  howpublished={SGLang Documentation, \url{https://docs.sglang.io/docs/advanced_features/hicache}},
  year={2025}
}

@inproceedings{ktransformers,
  title={{KTransformers: Unleashing the Full Potential of CPU/GPU Hybrid Inference for MoE Models}},
  author={Hongtao Chen and Weiyu Xie and Boxin Zhang and Jingqi Tang and Jiahao Wang and Jianwei Dong and Shaoyuan Chen and Ziwei Yuan and Chen Lin and Chengyu Qiu and Yuening Zhu and Qingliang Ou and Jiaqi Liao and Xianglin Chen and Zhiyuan Ai and Yongwei Wu and Mingxing Zhang},
  booktitle={ACM Symposium on Operating Systems Principles (SOSP)},
  year={2025}
}

@inproceedings{eagle3,
  title={{EAGLE-3: Scaling up Inference Acceleration of Large Language Models via Training-Time Test}},
  author={Yuhui Li and Fangyun Wei and Chao Zhang and Hongyang Zhang},
  booktitle={Advances in Neural Information Processing Systems (NeurIPS)},
  year={2025}
}

@inproceedings{longspec,
  title={{LongSpec: Long-Context Lossless Speculative Decoding with Efficient Drafting and Verification}},
  author={Penghui Yang and Cunxiao Du and Fengzhuo Zhang and Haonan Wang and Tianyu Pang and Chao Du and Bo An},
  booktitle={Annual Meeting of the Association for Computational Linguistics (ACL)},
  year={2026}
}

@article{owl,
  title={{OWL: Overcoming Window Length-Dependence in Speculative Decoding for Long-Context Inputs}},
  author={Jaeseong Lee and {Seung-won} Hwang and Aurick Qiao and Gabriele Oliaro and Ye Wang and Samyam Rajbhandari},
  journal={arXiv preprint arXiv:2510.07535},
  year={2025}
}

@article{llama3,
  title={{The Llama 3 Herd of Models}},
  author={Aaron Grattafiori and Abhimanyu Dubey and others},
  journal={arXiv preprint arXiv:2407.21783},
  year={2024}
}

@article{onlinesoftmax,
  title={{Online normalizer calculation for softmax}},
  author={Milakov, Maxim and Gimelshein, Natalia},
  journal={arXiv preprint arXiv:1805.02867},
  year={2018}
}

@inproceedings{flashattn2,
  title={{FlashAttention-2: Faster Attention with Better Parallelism and Work Partitioning}},
  author={Dao, Tri},
  booktitle={International Conference on Learning Representations (ICLR)},
  year={2024}
}

@inproceedings{gqa,
  title={{GQA: Training Generalized Multi-Query Transformer Models from Multi-Head Checkpoints}},
  author={Joshua Ainslie and James Lee-Thorp and Michiel de Jong and Yury Zemlyanskiy and Federico Lebr{\'o}n and Sumit Sanghai},
  booktitle={Conference on Empirical Methods in Natural Language Processing (EMNLP)},
  year={2023}
}

@inproceedings{quest,
  title={{Quest: Query-Aware Sparsity for Efficient Long-Context LLM Inference}},
  author={Jiaming Tang and Yilong Zhao and Kan Zhu and Guangxuan Xiao and Baris Kasikci and Song Han},
  booktitle={International Conference on Machine Learning (ICML)},
  year={2024}
}

@article{roofline,
  title={{Roofline: An Insightful Visual Performance Model for Multicore Architectures}},
  author={Williams, Samuel and Waterman, Andrew and Patterson, David},
  journal={Communications of the ACM},
  volume={52},
  number={4},
  pages={65--76},
  year={2009}
}

@inproceedings{lwm,
  title={{World Model on Million-Length Video and Language with Blockwise RingAttention}},
  author={Hao Liu and Wilson Yan and Matei Zaharia and Pieter Abbeel},
  booktitle={International Conference on Learning Representations (ICLR)},
  year={2025}
}

@inproceedings{beyondgpu,
  title={{Scaling Attention Beyond GPUs for LLM Inference}},
  author={Weishu Deng and Yujie Yang and Peiran Du and Lingfeng Xiang and Zhen Lin and Chen Zhong and Faraz Ahmed and Lianjie Cao and Puneet Sharma and Song Jiang and Hui Lu and Jia Rao},
  booktitle={International Symposium on High-Performance Parallel and Distributed Computing (HPDC)},
  year={2026}
}

@article{dhisparse,
  title={{Adaptive DHiSparse: Speculative Decoding over a Load-Adaptive KV Hierarchy}},
  author={Naji Alazhar and Wassel Alazhar},
  journal={arXiv preprint arXiv:2610.13708},
  year={2026}
}

@inproceedings{sparsespec,
  title={{Accelerating Large-Scale Reasoning Model Inference with Sparse Self-Speculative Decoding}},
  author={Yilong Zhao and Jiaming Tang and Kan Zhu and Zihao Ye and Chi-Chih Chang and Chaofan Lin and Jongseok Park and Guangxuan Xiao and Mohamed S. Abdelfattah and Mingyu Gao and Baris Kasikci and Song Han and Ion Stoica},
  booktitle={Conference on Machine Learning and Systems (MLSys)},
  year={2026}
}

@inproceedings{directkv,
  title={{No Buffer, No Bottleneck: Efficient Zero-Copy KV Cache Offloading for Long-Context LLMs}},
  author={Shutian Luo and Haiying Shen},
  booktitle={USENIX Symposium on Operating Systems Design and Implementation (OSDI)},
  year={2026}
}

@inproceedings{echo,
  title={{ECHO: Efficient KV Cache Offloading with Lossless Prefetching for Serving Native Sparse Attention LLMs}},
  author={Guangda Liu and Wenhao Chen and Chengwei Li and Zhenyu Ning and Jing Lin and Yiwu Yao and Quan Chen and Shixuan Sun and Jieru Zhao and Minyi Guo},
  booktitle={USENIX Symposium on Operating Systems Design and Implementation (OSDI)},
  year={2026}
}

@misc{specstreamrepo,
  author={{LfLab-AI}},
  title={{SpecStream (SGLang fork)}},
  howpublished={\url{https://github.com/LfLab-AI/SpecStream}, commit e0b82d5},
  year={2026}
}

@inproceedings{hftransformers,
  title={{Transformers: State-of-the-Art Natural Language Processing}},
  author={Thomas Wolf and Lysandre Debut and Victor Sanh and Julien Chaumond and Clement Delangue and Anthony Moi and Pierric Cistac and Tim Rault and R{\'e}mi Louf and Morgan Funtowicz and Joe Davison and Sam Shleifer and Patrick von Platen and Clara Ma and Yacine Jernite and Julien Plu and Canwen Xu and Teven Le Scao and Sylvain Gugger and Mariama Drame and Quentin Lhoest and Alexander M. Rush},
  booktitle={Conference on Empirical Methods in Natural Language Processing: System Demonstrations (EMNLP)},
  year={2020}
}

@misc{llama32,
  title={{Llama 3.2: Revolutionizing Edge AI and Vision with Open, Customizable Models}},
  author={{Meta AI}},
  howpublished={\url{https://ai.meta.com/blog/llama-3-2-connect-2024-vision-edge-mobile-devices/}},
  year={2024}
}

@article{hybridgen,
  title={{HybridGen: Efficient LLM Generative Inference via CPU-GPU Hybrid Computing}},
  author={Mao Lin and Xi Wang and Guilherme Cox and Dong Li and Hyeran Jeon},
  journal={arXiv preprint arXiv:2604.18529},
  year={2026}
}

\end{document}